\documentclass[draft]{agujournal2019}
\usepackage{url} 
\usepackage{lineno}
\usepackage[inline]{trackchanges} 
\usepackage{soul}
\usepackage{amsmath}
\usepackage{amsfonts}
\usepackage{amssymb}
\draftfalse

\journalname{JGR: Space Physics}

\begin{document}

%
%


\title{Extreme, transient bursts of energy in the auroral ionosphere. II. A magnetotail dipolarization event}


%
%




\authors{Magnus F Ivarsen\affil{1,2}\thanks{Contact: magnus.fagernes@gmail.com}, Yukinaga Miyashita\affil{3,4}, Brian Pitzel\affil{1}, Jean-Pierre St-Maurice\affil{1,5}, Jaeheung Park\affil{3,4}, Devin R Huyghebaert\affil{6,1}, Yangyang Shen\affil{7}, and Glenn C Hussey\affil{1}}

\affiliation{1}{Department of Physics and Engineering Physics, University of Saskatchewan, Saskatoon, Canada}
\affiliation{2}{Department of Physics, University of Oslo, Oslo, Norway}
\affiliation{3}{Center for Heliophysics Research, Korea Astronomy and Space Science Institute, Daejeon, South Korea}
\affiliation{4}{Department of Astronomy and Space Science, Korea University of Science and Technology, Daejeon, South Korea}
\affiliation{5}{Department of Physics and Astronomy, University of Western Ontario, London, Canada}
\affiliation{6}{Leibniz Institute of Atmospheric Physics, K\"{u}hlungsborn, Germany}
\affiliation{7}{Department of Physics, University of Alberta, Edmonton, Canada}

\correspondingauthor{Magnus F Ivarsen}{magnus.fagernes@gmail.com}



\begin{keypoints}
\item \textsc{icebear} tracked auroral E-region radar targets with speeds exceeding 5000~m/s, implying transient electric fields that exceed 250~mV/m.
\item Transients were caused by magnetotail dipolarization, observed \textit{in-situ} by THEMIS, and Swarm detected Alfv\'{e}n waves on adjacent field lines.
\item Simple WKB wave-transmission analysis recovers the 250~mV/m electric field amplitude at the ionospheric foot via Alfv\'{e}n waves.
\end{keypoints}

%
%

%
%


\begin{abstract}
We report ground-based coherent VHF radar observations of extreme ($>5000$~m/s, mostly eastward) turbulent field-structures detected in coincidence with a magnetospheric substorm-associated dipolarization at geocentric distance $\approx 7-9R_E$. The field-structures are observed by the  \textsc{icebear} radar over Saskatchewan, Canada, in the form of Farley-Buneman (FB) waves in the auroral electrojets, and the field-structures themselves move an order of magnitude faster than the saturation speed of the underlying FB waves, implying transient electric field sources of $\sim330$~mV/m in strength, occurring mostly at the poleward auroral arc boundaries. The field-structures are identified and automatically tracked using an unsupervised clustering \& tracking algorithm, applied to clutters of \textsc{icebear} radar backscatter targets, a method that turns the Doppler radar into a \textit{tracking radar} capable of measuring the ionospheric $\mathbf{E}\times\mathbf{B}$-drift by proxy. We place this finding in a coordinated multi-instrument context. Three \textsc{themis} spacecraft (from the Time History of Events and Macroscale Interactions during Substorms mission), observed the dipolarization event \textit{in-situ} in the near-Earth plasma sheet. In the ionosphere, Swarm~A, crossing through the guilty auroral arc at the onset of the dipolarization event, recorded clear signatures of propagating Alfv\'{e}n waves threading the relevant flux tube. We interpret the \textsc{icebear} transients as the natural ionospheric foot signature of a shear Alfv\'{e}n pulse launched by the bipolar space-charge (Hall) electric field of the thinned current sheet, with amplification along the converging flux tube, partial reflection at the ionospheric boundary, and spatial sharpening by precipitation-produced Pedersen-conductance gradients on the auroral arc edges. A one-dimensional wave-transmission analysis recovers the observed $\sim330$~mV/m amplitude. Our results elucidate a tightly controlled coupling between magnetotail processes and meter-scale auroral plasma turbulence, and demonstrate the capability of the interferometric, ground-based coherent radar \textsc{icebear} to resolve extreme, transient electric-field enhancements in the ionosphere.
\end{abstract}


%
%

%


%
%
%
%
\section{Introduction}

The ionosphere's high-latitude electric field is set, to first order, by magnetospheric convection. Under quiet solar-wind conditions the convection electric field is typically below $\sim20$~mV/m in the ionosphere, organized into the canonical twin-cell pattern by the Dungey reconnection cycle \citep{dungeyInterplanetaryMagneticField1961,cowleyTUTORIALMagnetosphereIonosphereInteractions2000}, and routinely monitored by the SuperDARN HF radar network \citep{thomasStatisticalPatternsIonospheric2018}. Active solar-wind driving, particularly with southward interplanetary magnetic field (IMF), qualitatively changes this picture: the dayside magnetosphere is eroded, the magnetotail stores magnetic flux through stretching, and the night-side field are more stretched. When the stretched configuration becomes unstable, the field reconfigures abruptly toward a dipole-like state. This rapid reconfiguration, termed dipolarization, is a macroscopic signature of substorm onset \citep{ohtani_earthward_1998,ohtani_substorm_2001}.

The ionospheric consequences of dipolarization are quantitatively distinct from those of quiet-time convection. Dipolarization-associated transients launch shear Alfv\'{e}n-wave trains earthward along the converging auroral flux tube \citep{hongIonDynamicsAssociated2008,daiGeoeffectivenessInterplanetaryAlfven2023}, delivering concentrated electromagnetic energy to the upper atmosphere \citep{keilingGlobalMorphologyWave2003,pakhotinDiagnosingRoleAlfven2018}. The waves themselves are routinely observed \textit{in-situ} by satellites, but a single polar-orbiting, low-altitude satellite pass samples a narrow track and cannot resolve the lateral extent or temporal evolution of the disturbance at the field-line's footprint.

Ground-based coherent VHF radar offers a complementary observable. Wherever and whenever the ionospheric electric field exceeds $\sim20$ mV/m, meter-scale Farley-Buneman (FB) plasma turbulence is typically excited in the auroral electrojets at altitudes near 105~km \citep{farleyPlasmaInstabilityResulting1963,bunemanExcitationFieldAligned1963,sahr_auroral_1996}. For dipolarization-associated transients this threshold is comfortably exceeded, and coherent radars, given a sufficiently high spatio-temporal resolution, therefore act as wide-field, high-cadence detectors of the ionospheric foot of the magnetospheric driver \citep{ivarsen_eastward_2025,ivarsen_eastward_2025-1}. The  radar \textsc{icebear} \citep[Ionospheric Continuous wave E-region Bistatic Experimental Auroral Radar; ][]{huyghebaertICEBEARAlldigitalBistatic2019} provides 1-s temporal resolution and $\sim1$-km spatial resolution through a long interferometric baseline; compared to a conventional SuperDARN radar, this is a massive improvement in resolution of $60\times$ and $40\times$, respectively. Combined with the point-cloud clustering \& tracking method of \cite{ivarsen_point-cloud_2024,ivarsen_deriving_2024}, discrete FB source regions can be identified and their  target motions tracked individually, distinct from the underlying FB Doppler phase speeds, which are saturated near the local ion acoustic speed \citep{st.-mauriceNewNonlinearApproach2001,fosterSimultaneousObservationsEregion2000,oppenheim_kinetic_2013}.

\begin{figure*}
    \centering
    \includegraphics[width=0.78\textwidth]{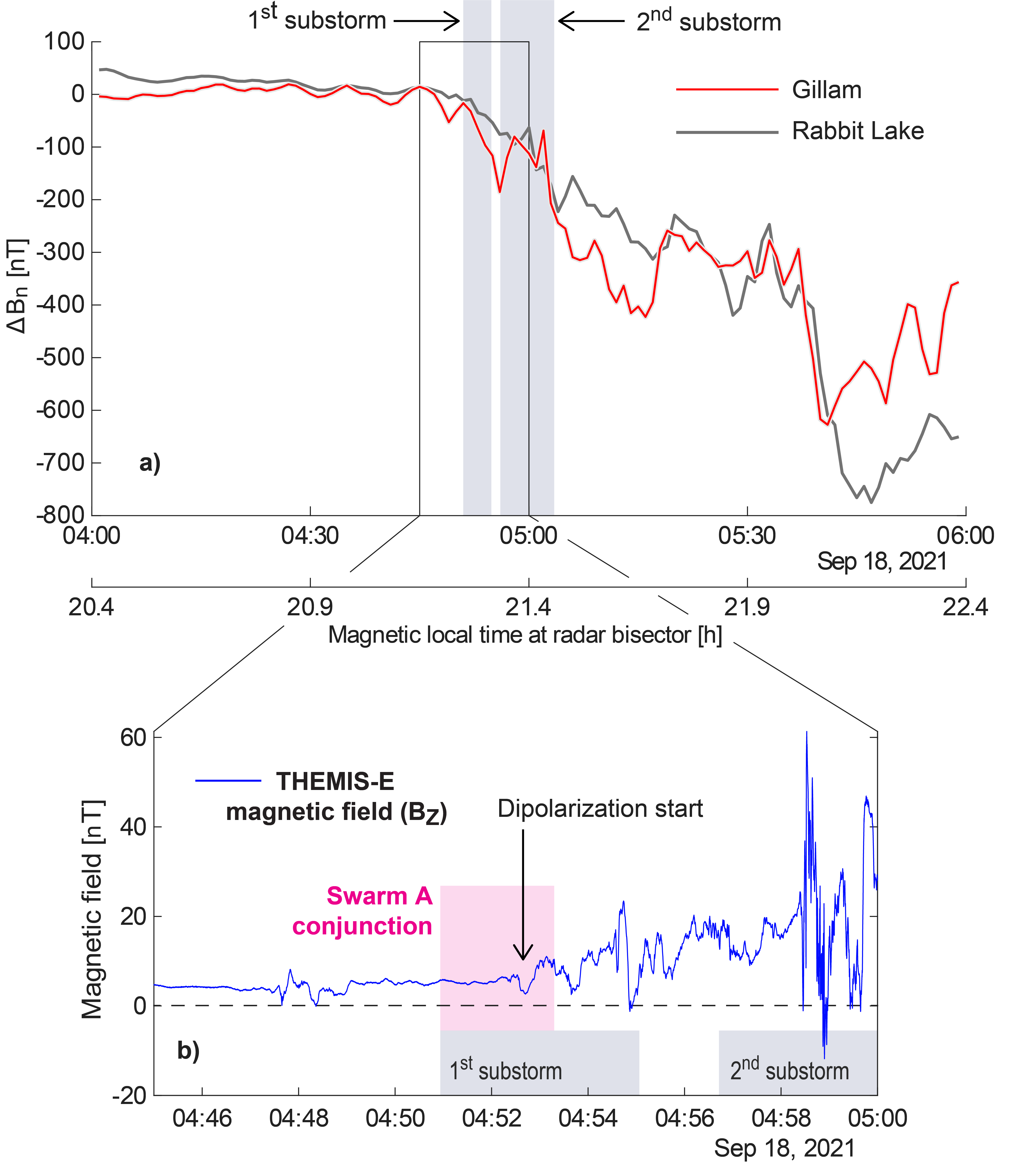}
    \caption{\textbf{Panel a):} The development of the auroral electrojets, or the high-latitude Hall currents, measured by ground-based magnetometers at Gillam (red line) and Rabbit Lake (grey line). Indicated are two separate substorms identified by examination of the auroral images (Figure~\ref{fig:extgillam}). See Figure~\ref{fig:imf} in Appendix~B for observations of the interplanetary magnetic field prior to the event under study. \textbf{Panel b):} \textit{In-situ} observations of the magnetic field $\mathbf{B}_Z$ by \textsc{themis}-E. See Figures~\ref{fig:extthem}, \ref{fig:themis} for details of the \textsc{themis} data.}
    \label{fig:opptakt}
\end{figure*}

In this paper we apply that capability to a multi-instrument conjunction over Saskatchewan, Canada, on 18 September 2021, following weak and moderate substorms with negative excursions of the northward magnetic field of ~-100 and -200 nT, respectively. Concurrently with the first substorm onset, a magnetotail dipolarization event was observed \textit{in-situ} by three \textsc{themis} spacecraft at $L \approx 7-9 R_E$ (Figure~\ref{fig:opptakt}b), and Swarm observed Alfv\'{e}n waves at the field-line's foot. Concurrent with dipolarization with rapid large fluctuations associated with the second substorm, \textsc{icebear} tracked extremely fast E-region transients (up to $6000$~m/s) on the poleward edges of the developing auroral forms, implying exceedingly strong transient transverse electric fields of up to $\sim330$ mV/m.

\begin{figure}
    \centering
    \includegraphics[width=.63\textwidth]{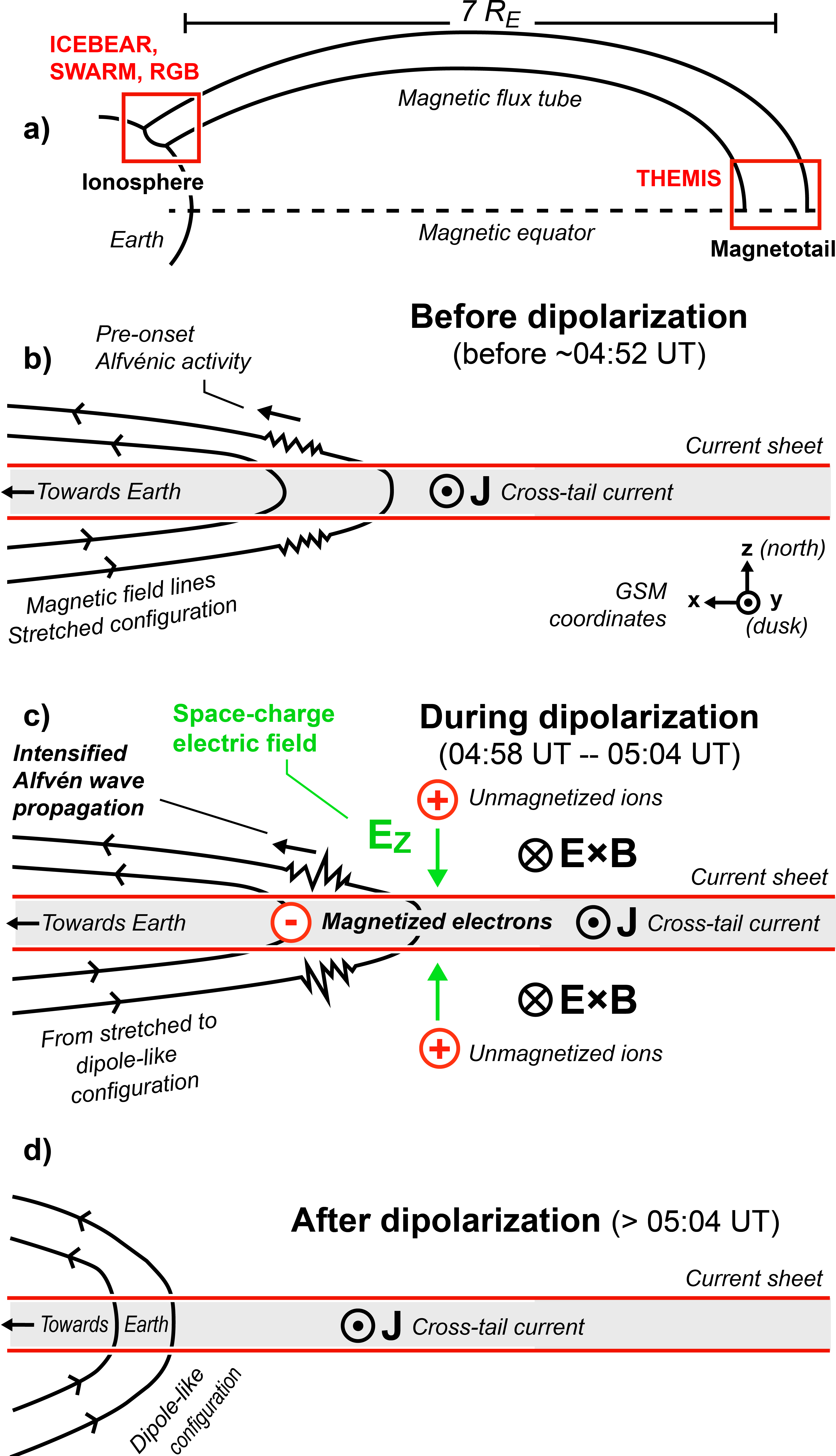}
    \caption{\textbf{Panel a):} Magnetic flux tube connecting the auroral ionosphere over Saskatchewan, Canada, to the near-Earth plasma sheet at $X \approx -7 R_E$, with the magnetic equator indicated. \textbf{Panel b):} Pre-onset phase (before 04:52~UT), exhibiting a stretched configuration with intact cross-tail current sheet (J duskward) and pre-existing Alfv\'{e}nic activity threading the flux tube. \textbf{Panel c):} Dipolarization with large fluctuations at $\sim$04:58-05:04~UT: thinning of the current sheet to ion-gyroradius scales decouples ions from $\mathbf{B}$ while electrons remain magnetized, producing a bipolar space-charge (Hall) electric field $\mathbf{E}_Z$ directed \textit{toward the central current sheet from both sides} \citep{luHallElectricField2019}. The associated dawnward electron Hall drift, $\mathbf{v}_e = \mathbf{E}_Z\times\mathbf{B}_X / B^2$, sustains the duskward cross-tail current ($\mathbf{J} = -en_e\mathbf{v}_e$) locally, and the transient $\mathbf{E}_Z$ launches an intensified shear Alfv\'{e}n pulse earthward along the converging flux tube (see Appendix~A). \textbf{Panel d):} Post-dipolarization ($\sim$05:04~UT): the current sheet thickens, ions re-magnetize, the bipolar $\mathbf{E}_Z$ disappears, and the field conventionally returns to a dipole-like configuration. Coordinates (geocentric solar magnetospheric, GSM) in panels b--d) are right-handed with $x$ earthward ($+X_\text{GSM}$), $z$ northward ($+Z_\text{GSM}$), and $y$ duskward ($+Y_\text{GSM}$). 
    }
    \label{fig:cartoon}
\end{figure}

\section{Methodology \& theoretical foundation}

This study uses measurements from ground-based optical and radar systems, a low-Earth orbit satellite, and near-equatorial magnetospheric spacecraft. Together, these instruments capture an electrodynamic sequence initiated by a substorm-associated dipolarization and earthward flow burst at roughly $7R_E$, and its electrodynamic consequences in the upper atmosphere. In this section, we outline the theoretical framework linking the observed magnetotail dynamics to E-region ionospheric turbulence observations. 

The interval of interest occurred under moderately active conditions, characterized by a relatively weak background convection electric field, immediately preceding a series of weak and moderate substorms (Figure~\ref{fig:opptakt}a), and we note that this magnetospheric event was recently studied by \cite{babu_plasma_2024}.

\subsection{Magnetospheric measurements} \label{sec:theory}

Figure \ref{fig:cartoon} illustrates the anticipated electrodynamic coupling along a magnetic flux tube connecting the magnetotail to the nightside ionosphere during such an event. In the substorm expansion phase, the disruption of the cross-tail current leads to a rapid reconfiguration of the local magnetic field from a stretched, tail-like topology to a dipole-like configuration \citep[e.g., ][]{ohtani_earthward_1998,ohtani_substorm_2001}. This dipolarization process is typically accompanied by highly structured magnetic fields \citep{ohtani_substorm_2001} and drives energetic particle precipitation into the nightside ionosphere \citep{kabin_particle_2017,duan_characteristics_2021}.

As illustrated in Figure \ref{fig:cartoon}b, in the thin current sheet where dipolarization occurs, ions become unmagnetized while electrons remain magnetized, causing charge separation and hence equatorward electric fields (Hall $\boldsymbol{E_Z}$) above and below the equator \citep{luHallElectricField2019}. This transient electric field serves as the source condition for an Alfv\'{e}nic pulse that propagates earthward along the magnetic field lines toward ionospheric altitudes \citep{hongIonDynamicsAssociated2008}, and is expected to decay over the course of a few minutes as the source structure diffuses.

Simultaneously, in the ionosphere, the flux tube supports a current of precipitating electrons, forming the visible auroral arc \citep{rothTheoreticalStructureMagnetospheric1993,borovskyQuiescentDiscreteAuroral2019,lysakQuietDiscreteAuroral2020}. These electrons are accelerated downward by wave-particle interactions \citep{chaston_turbulent_2008,shen_red_2024} or parallel potential drops \citep{echimMagnetosphericGeneratorDriving2009}, often in association with kinetic Alfv\'{e}n waves, ultimately triggered by the turbulent magnetotail structuring. The field modulations are expected to produce equatorward electric fields on the poleward edge of the auroral structures \citep{opgenoorthRegionsStronglyEnhanced1990,hosokawaLargeFlowShears2013,gallardo-lacourt_ionospheric_2014,ivarsen_turbulence_2024}.

The transient space-charge field ($\boldsymbol{E_Z}$) at the dipolarization thereby sets the amplitude of an Alfv\'{e}n pulse whose perpendicular electric field evolves along the converging flux tube according to the local Alfv\'{e}n-speed and magnetic-field profile, and whose amplitude at the ionospheric foot is determined by partial reflection against the boundary impedance set by the ratio of the Alfv\'{e}n-wave admittance to the height-integrated Pedersen conductance. The transmitted field is further spatially concentrated at the foot by the localized conductance structure of the auroral arc, where the polarization electric fields produced by Cowling channels on the edges of auroral forms \citep{fujii_reformulation_2011,fujii_application_2012} should sharpen the electrodynamic response. We hypothesize that this wave-transmitted, conductance-modulated field produces the transient, extreme electric field enhancements required to drive the super-fast, radar-tracked E-region field-structures observed by the coherent scatter radar in this study.

\begin{figure}
    \centering
    \includegraphics[width=1.1\textwidth]{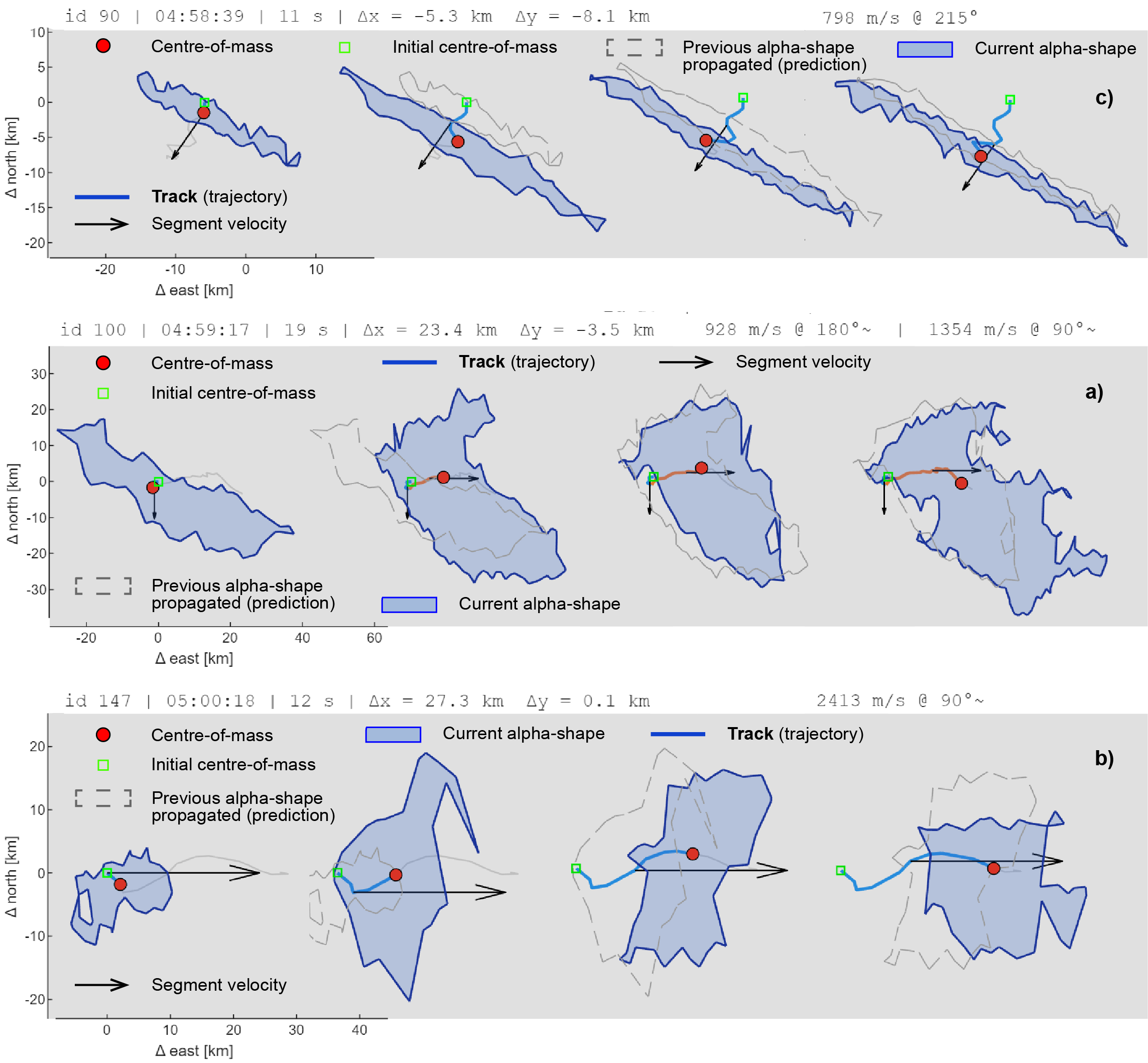}
\caption{Three examples of tracked echo clusters observed on 18 September 2021, with displacement and speed posted above each line. The evolution of each cluster is shown in four equally spaced temporal snapshots. The clusters are represented by their alpha-shape (blue shaded region), enclosing the echo point-clouds in polygons. For each step, the previous alpha-shape is shown with dashed lines, and a red circle indicates the echo centre-of-mass; a green square indicates the initial echo centre-of-mass. The tracks, or cluster trajectories, are plotted for each segment of the cluster's spatio-temporal evolution. Arrows indicate the directions of travel in those segments. 
}
    \label{fig:icebear}
\end{figure}

To elaborate on this ``\textit{mapping}'' of a magnetotail dipolarization event
from the near-Earth magnetotail down to the auroral E-region, we must comment on
Wentzel–Kramers–Brillouin (WKB) propagation along the converging flux tube. Its
action amplifies the perpendicular electric field as $E_\perp\propto\sqrt{v_A B}$
\citep{lysakFeedbackInstabilityIonospheric1991}, $v_A$ being the Alfv\'{e}n speed,
giving an amplification of $\approx25$--$57$ between the near-Earth plasma
sheet and the auroral field-line's foot along the three field lines occupied by three \textsc{themis} spacecraft
(Eq.~\ref{eq:ampfoot}; Table~\ref{tab:wkb}). Partial reflection at the ionospheric
boundary then occurs, governed by the ratio of the Alfv\'{e}n-wave admittance
$\Sigma_A=1/(\mu_0 v_A)$ to the height-integrated Pedersen conductance $\Sigma_P$
\citep{yoshikawaReflectionShearAlfven1996,parkAlfvenWavesAuroral2017,ivarsenObservationalEvidenceRole2020}.
With a source field of $20$--$50$~mV\,m$^{-1}$ this yields a foot field of
$\sim$80--700~mV\,m$^{-1}$ for the nominal (\textsc{themis}-D) field line, comfortably bracketing
the extreme $\sim$330~mV\,m$^{-1}$ inferred from the eastward \textsc{icebear} irregularity
$\mathbf{E}\times\mathbf{B}$ drifts. See Appendix~A for our rudimentary simulation
treatment of this Alfv\'{e}nic mapping.

\subsection{Ionospheric measurements}

The turbulence that tracks extreme electric field spikes, observations of which comprise the main results of the present study, are a multitude of intense, meter-scale turbulent waves referred to as Farley-Buneman (FB) waves, excited inside the auroral electrojets (Hall currents). Each wave is the result of a modified two-stream fluid instability mechanism \citep{farleyPlasmaInstabilityResulting1963,bunemanExcitationFieldAligned1963}. Localized increases in the electric fields introduce a differential motion between electrons and ions, exciting unstable plasma waves with growing amplitudes that move in the $\boldsymbol{E}\times\boldsymbol{B}$-direction (Hall drifts), with $\boldsymbol{E}$ and $\boldsymbol{B}$ being the electric and magnetic fields. 

FB waves scatter incident radar signals, and are thus the source of the radar aurora. That is, plasma turbulence capable of causing Bragg-like scattering of radar signals. As mentioned, these turbulent regions of the ionospheric plasma tend to hug the outlines of auroral arcs \citep{ivarsen_turbulence_2024}, at a distance from the arcs' external boundaries \citep{bahcivanObservationsColocatedOptical2006,huyghebaertPropertiesICEBEARERegion2021}.
Notably, a new method presented in \cite{ivarsen_point-cloud_2024} and \cite{ivarsen_deriving_2024} allows for the accurate determination of the $\mathbf{E}\times\mathbf{B}$-drift speeds from coherent E-region radar alone. These are the \textsc{icebear} \textit{target motions}, which are distinguished from the traditional radar Doppler speeds. The latter are saturated by the ion acoustic (sound) speed, and cannot yield information about the electric field without considerable scrutiny \citep{hysellComparingVHFCoherent2012}. The automatic tracking of these apparent motion of echo structures, turning the Doppler radar \textsc{icebear} into a tracking radar, constitutes a proxy measurement for the $\mathbf{E}\times\mathbf{B}$-drift speed. A sparse, yet growing, body of literature verifies the method's efficacy \citep{ivarsen_point-cloud_2024,ivarsen_deriving_2024,ivarsen_eastward_2025-1}.

In this study, we implement an improved variant of the tracking algorithm presented in \cite{ivarsen_point-cloud_2024}. The improvements entail efficient representation of the clustered point-clouds as `alpha-shapes' \citep{edelsbrunnerThreedimensionalAlphaShapes1994} as the per-frame detector and the Hungarian algorithm \citep{kuhnHungarianMethodAssignment1955} for as a linear-assignment solver under a centroid-distance plus Intersection-over-Union (IoU) cost as the cross-frame associator  \citep{bewleySimpleOnlineRealtime2016,wojkeSimpleOnlineRealtime2017}, regularized by a constant-velocity predictor that tightens the gating radius once a track has an estimated velocity. A degenerate Kalman filter allows for monitoring of the tracked tragets \citep{yangExtendedKalmanFilter2017}, and a sequential piecewise-linear regression sliding-window segmentation \citep{keoghOnlineAlgorithmSegmenting2001} yield accurate target velocities. This improved radar tracking method is presented in detail in a companion manuscript \citep{ivarsenPredictiveRadarTracking2026}.

Figure~\ref{fig:icebear} shows three particular \textsc{icebear} echo clusters, observed, identified, and tracked during the magnetotail dipolarization event under study. 

\section{\label{sec:data}Data}

In this section, we shall present a comprehensive description and summary of the data we analyze in the present paper.

\subsection{Auroral observations} \label{sec:aurora}

\begin{figure}
    \centering
    \includegraphics[width=1.1\textwidth]{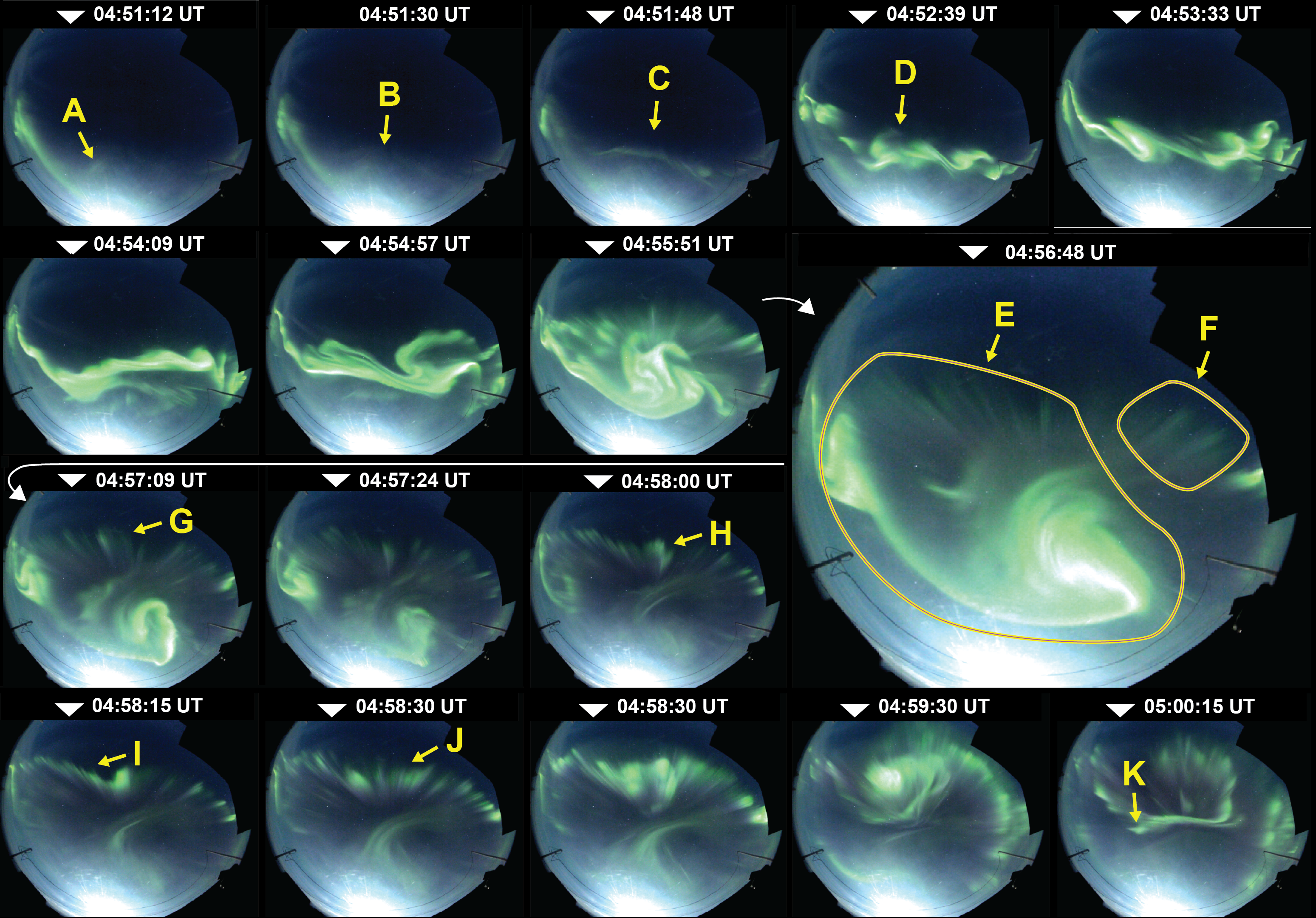}
    \caption{Optical images by \textsc{tre}x \textsc{rgb} at Gillam ($56.4^\circ$~N, $94.7^\circ$~W); geographic north is up, west is to the left; the white patch towards the bottom of the frame is the Moon. \textbf{Annotations:} \textbf{A:} Initial brightening (at first, very faint). \textbf{B:} Gradually growing and extending. \textbf{C:} Further enhancement of the onset arc. \textbf{D:} Enhanced wavelike (bead-like) structure. \textbf{E:} The first substorm is still subsiding. \textbf{F:} Initial brightening of the second substorm (new activity), poleward of the auroral activity of the first substorm. \textbf{G:} Gradually growing and extending. \textbf{H:} Further enhancement of the onset arc. \textbf{I:} Enhanced wavelike (bead-like) structure. \textbf{J:} Poleward expansion. \textbf{K:} Another poleward expansion.
}
    \label{fig:extgillam}
\end{figure}

We begin by showing a series of auroral images captured by the \textsc{tre}x \textsc{rgb} system of auroral ``all-sky-imagers'' (ASIs), which we map to a fixed altitude (105~km) \citep{gilliesApparentMotionSTEVE2020,trex_rgb_ucalgary_2023}. The imagers are located at the Rabbit Lake ($58.2^\circ$~N, $103.7^\circ$~W in geographic coordinates) and Gillam ($56.4^\circ$~N, $94.7^\circ$~W) research stations in Canada. 

Figure~\ref{fig:extgillam} shows selected 3-second auroral images from Gillam to show the full sequence of auroral development in the interval of interest (see also Movie S1 in the supporting information). In this interval, a pseudobreakup (pseudosubstorm) and two substorms occurred in the camera's field of view. The auroral onset arcs developed stepwise, akin to descriptions of substorms by \cite{miyashitaRevisitingSubstormEvents2018} and \cite{miyashita_case_2018,miyashitaEvolutionNearEarthMagnetotail2025}.

First, initial auroral brightening of the pseudobreakup began near the southwestern edge of the field of view at $\sim$04:46:45 UT, and the initial brightening (auroral onset) arc gradually grew and extended eastward (see Movie~S1). Then, it further enhanced near the southern edge of the field of view at $\sim$04:47:42 UT, and somewhat large wavelike (bead-like) structure appeared. It, however, subsided without progressing to poleward expansion (auroral breakup), so this auroral activity is regarded as a pseudobreakup \citep{miyashita_case_2018}.

Soon after that, initial auroral brightening of the first substorm began poleward of the pseudobreakup arc's location and to the south of the zenith at $\sim$04:51:12 UT (Figure~\ref{fig:extgillam}, A). The auroral onset arc was initially very faint, but it gradually grew and extended mainly eastward (Figure~\ref{fig:extgillam}, B). Then it further enhanced widely eastward of the initial brightening site at $\sim$04:51:48 UT (Figure~\ref{fig:extgillam}, C), and wavelike structure largely developed
into vortex-shaped forms. At $\sim$04:53:33 UT poleward expansion began in the eastern part of the onset arc (Figure~\ref{fig:extgillam}, D). The auroral breakup began to subside at $\sim$04:56 UT.

While the breakup aurora of the first substorm was still subsiding (Figure~\ref{fig:extgillam}, E), initial auroral brightening of the \textit{second} substorm began poleward of the first substorm's aurorae and to the east of the zenith at $\sim$04:56:48 UT (Figure~\ref{fig:extgillam}, F). The auroral onset arc then gradually grew and extended westward and eastward (Figure~\ref{fig:extgillam}, G). This is not a continuation of the first substorm but a new activity in a different region, although they were spatially and temporally close.

At $\sim$04:58:00 UT the onset arc further enhanced near the zenith (Figure~\ref{fig:extgillam}, H), and wavelike structure developed (Figure~\ref{fig:extgillam}, I). At $\sim$04:58:30 UT, poleward expansion began near the zenith (Figure~\ref{fig:extgillam}, J). For this substorm, another poleward expansion occurred to the south of the zenith at $\sim$05:00:15 UT (Figure~\ref{fig:extgillam}, K). Consistent with the poleward expansions associated with the two substorms, weak to moderate geomagnetic negative bays with $\sim$100-300 nT were observed at Gillam and nearby stations \citep[the CARISMA magnetometers, ][not shown]{mannUpgradedCARISMAMagnetometer2008}.

The auroral images in Figure~\ref{fig:extgillam} show a large degree of characteristic wavelike structure likely due to large-scale instabilities in the magnetotail, such as ballooning instability \citep{voronkov_coupling_1997, babu_plasma_2024}. Furthermore, of some interest, small-scale transient auroral beads appeared in the auroral arcs during auroral breakup, for example, at 04:56-04:57~UT. These structures may be caused by cross-field current instabilities as suggested by \cite{lui_cross-field_2016}, but their small-scale sizes of a few 10s of kilometers invite further scrutiny in a future study.

\begin{figure*}
    \centering
    \includegraphics[width=1.1\textwidth]{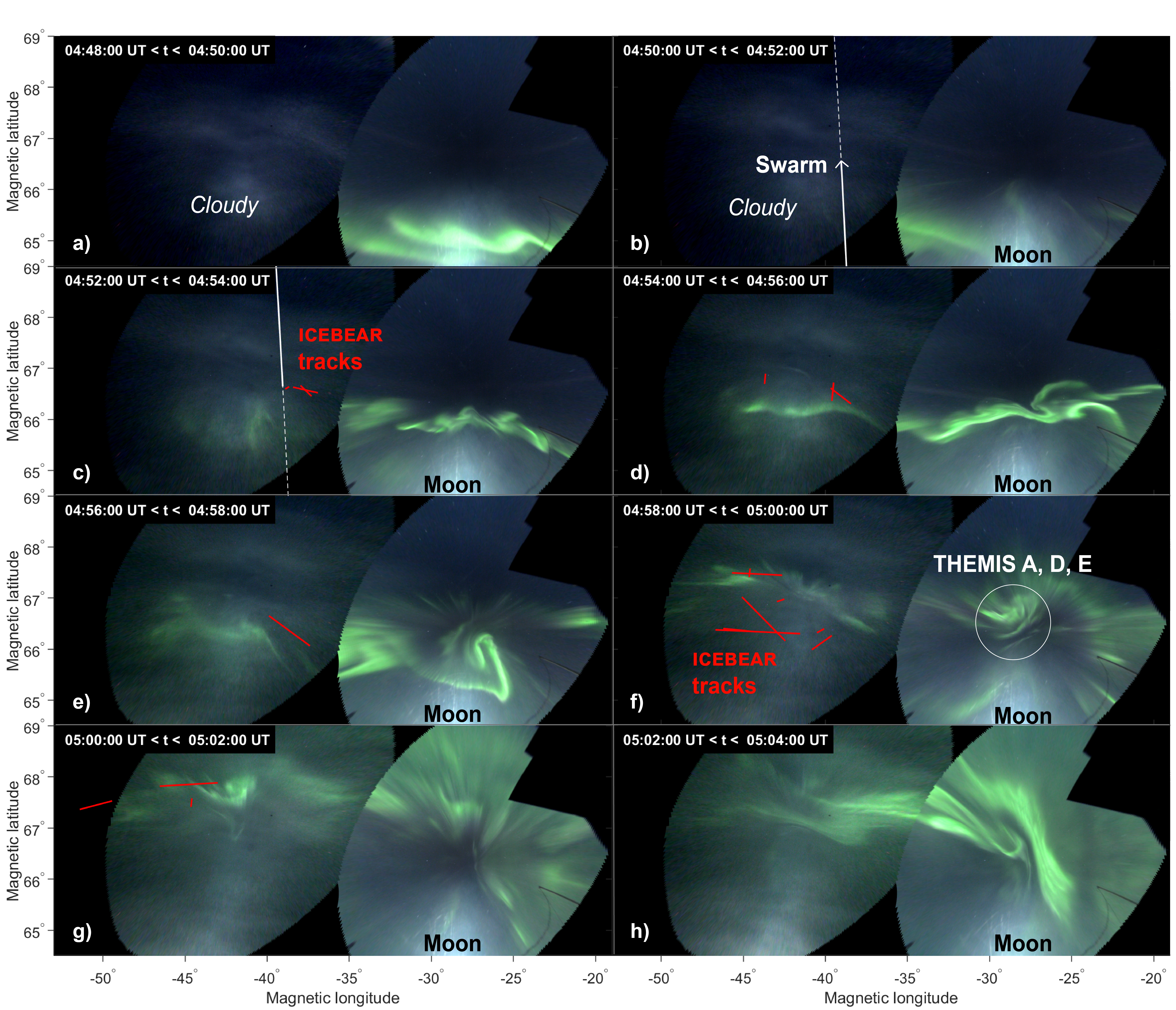}
    \caption{\textbf{A multiple conjunction event that took place between 04:48~UT and 05:04~UT on 18 September 2021.} Each panel (a--h) shows the superposition of two \textsc{tre}x \textsc{rgb} auroral cameras \citep{gilliesApparentMotionSTEVE2020}, with the stations Rabbit Lake to the left and Gillam to the right; the former saw some loose cloud cover throughout the event, and the latter had the moon within the field-of-view throughout the event. The trajectory of Swarm A is indicated in panels b) and c) (white line); the trajectories of tracked \textsc{icebear} echo clusters are shown with thin, red lines; the ionospheric footprint of \textsc{themis}~A, D, and E are indicated in Panel f). Magnetic longitude and latitude are indicated on the axes, and all data is shown with a field-line tracing coordinate system or altitude-adjusted corrected geomagnetic (AACGM) coordinates \citep{bakerNewMagneticCoordinate1989}. We note that close inspection of panel f) shows that the lower-latitude tracks appear on the poleward side of weaker, optical arcs.
    }
    \label{fig:synopsis}
\end{figure*}

Keeping the above-mentioned auroral development as a backdrop, we shall now describe the data from our three main sources of instrumentation, the radar \textsc{icebear}, the satellite Swarm~A, and the magnetospheric probes \textsc{themis}.
Figure~\ref{fig:synopsis} shows 3-second auroral snapshots that combine images from Gillam and Rabbit Lake, mapped to a geographic grid at an altitude of 105~km, and ranging from 04:48~UT to 05:04~UT. The binned intervals (2 minutes each) contain \textsc{icebear} (red tracks) and Swarm (white line) observations; the approximate ionospheric footprints of the \textsc{themis} spacecraft are shown in Panel f) (see Figures~\ref{fig:extthem}-\ref{fig:themis} below). They were located near the magnetic zenith of Gillam, separated by less than 1~h in magnetic local time (MLT) from the Swarm A trajectory and Rabbit Lake, and in the poleward part of the breakup aurora of the first substorm and near the auroral onset arc of the second substorm (see Figure~\ref{fig:extgillam}).

\subsection{\textsc{THEMIS} magnetotail observations} \label{sec:themis}

The \textsc{themis} mission \citep{angelopoulosTHEMISMission2008} utilizes a constellation of 5 identical probes in highly elliptical, near-equatorial orbits to pinpoint the macroscopic physical trigger of magnetospheric substorms.
In this study we used ion and magnetic and electric field data  from the \textsc{themis} A, D, and E spacecraft. Low- and high-energy ions were detected by the electrostatic analyzer \citep[ESA; ][]{mcfaddenTHEMISESAPlasma2008}  and the solid state telescope \citep[SST; ][]{angelopoulosTHEMISMission2008}, respectively, at 3 s resolution. Ion moments were calculated from the ESA and SST data. Magnetic fields were measured by the fluxgate magnetometer \citep[FGM; ][]{austerTHEMISFluxgateMagnetometer2008}   at 0.25 s resolution. Electric fields were measured by the electric field instrument \citep[EFI; ][]{bonnellElectricFieldInstrument2008}  at 0.25 s resolution. The component parallel to the spacecraft spin axis was estimated using the two components in the spacecraft spin plane under the assumption of ${\bf E}\cdot{\bf B}=0$, where the angle between the spin axis and the magnetic field should be $<$80$^\circ$. The data shown are in GSM coordinates.

\begin{figure}
    \centering
    \includegraphics[width=1.1\textwidth]{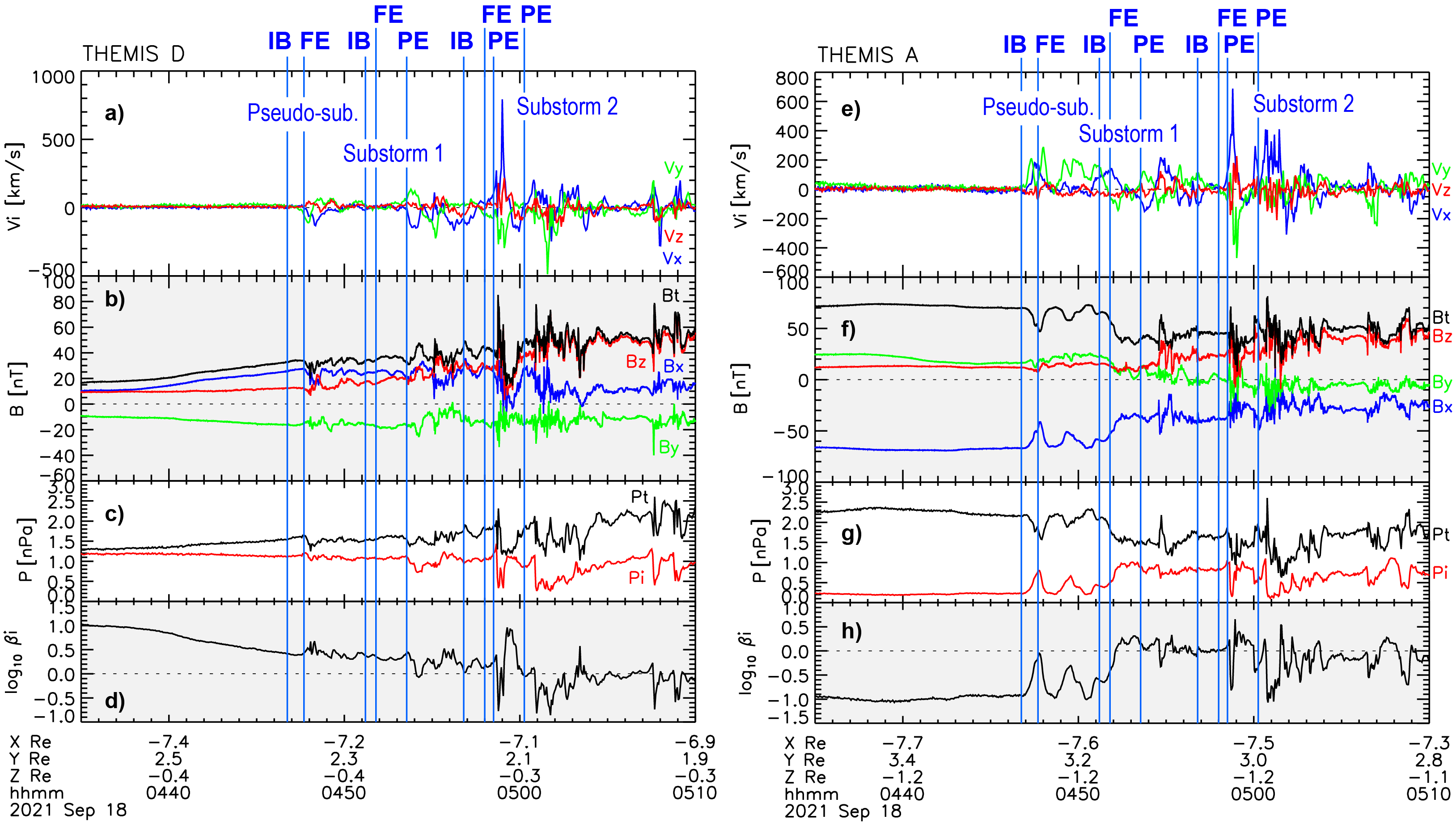}
    \caption{Ion velocity and the magnetic field in GSM coordinates, the total and ion pressures, and the ion $\beta$ from the \textsc{themis} D (left column, \textbf{panels a--d}), and E (right column, \textbf{panels e--h}) spacecraft from 04:35 to 05:10 UT on 18 September 2021. The total pressure (black line in panels c and g) is defined as the sum of the ion (red line in those panels) and magnetic pressures. The vertical lines from the left indicate \textbf{initial auroral brightening (IB)} and \textbf{further enhancement of the auroral onset arc (FE)} for the pseudobreakup; IB, FE, and auroral \textbf{poleward expansion (PE)} for the first substorm; and IB, FE, and PE for the second substorm.}   
    \label{fig:extthem}
\end{figure}

\begin{figure}
    \centering
    \includegraphics[width=0.63\textwidth]{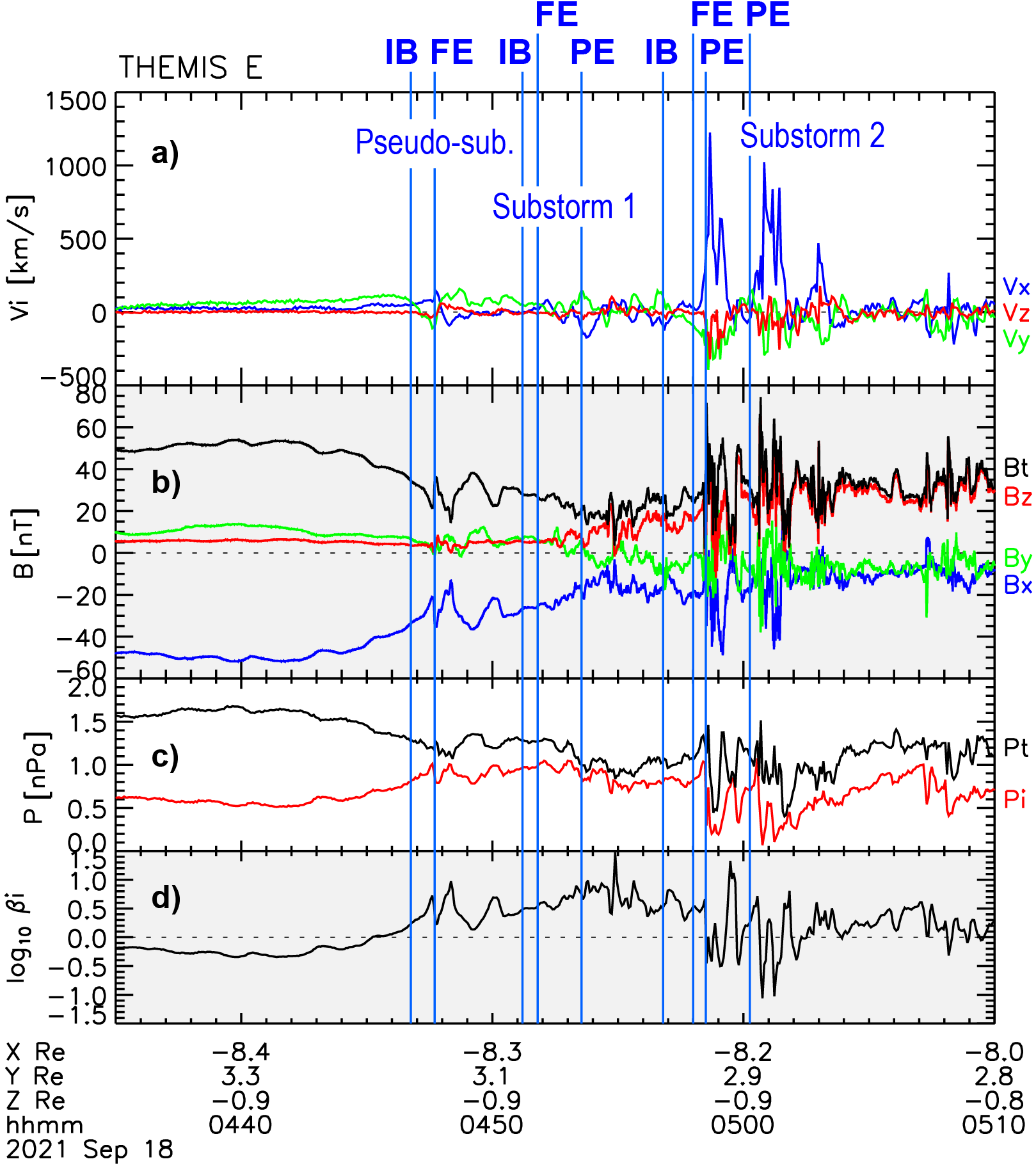}
    \caption{Ion velocity and the magnetic field in GSM coordinates, the total and ion pressures, and the ion $\beta$ from \textsc{themis} E. See Figure~\ref{fig:extthem} for detailed caption.}   
    \label{fig:extthem2}
\end{figure}

Figures~\ref{fig:extthem} and \ref{fig:extthem2} show the ion velocity, the magnetic field, the total and ion pressures, and the ion $\beta$ \textit{in-situ} measured by the three \textsc{themis} spacecraft at $X \sim -7$ to $-8$ $R_E$ and $Y \sim 2$ to 3 $R_E$ for the period from 04:35~UT to 05:30~UT. Until just before initial auroral brightening of the pseudobreakup, all three \textsc{themis} spacecraft registered magnetic field line stretching during the substorm growth phase, as indicated by increase in $|B_x|$. All three spacecraft were in the plasma sheet (see the ion pressure in Figures~\ref{fig:extthem} and \ref{fig:extthem2} and the particle spectrograms in Figures S1--S3 in supporting information), but \textsc{tha} was well south of the magnetic equator ($B_x \approx -65$~nT), featuring a small positive $B_z$ ($\sim$13~nT) and modest $B_y$ ($\sim$25~nT). \textsc{the} was situated less deeply south of the equator ($B_x \approx -45$~nT) with small $B_z$ and $B_y$, but it moved toward the equator at $\sim$04:45 UT, as indicated by decrease in $|B_x|$. \textsc{thd} was in the northern central plasma sheet with all three components below $\sim$15~nT.

At $\sim$04:47 UT, fluctuations in the Pi2 frequency range began, associated with the pseudobreakup. The amplitudes at \textsc{tha} and \textsc{the} were larger than that at \textsc{thd}. Weak fluctuations in the Pi1 frequency range also appeared during the pseudobreakup.

After that, dipolarization, persistent net increase in $B_z$, occurred twice, associated with the two substorms. Dipolarization associated with the first substorm began with slight increase in $B_z$ at \textsc{thd} and \textsc{the} at $\sim$04:51:40 and $\sim$04:52:00, respectively, nearly simultaneously with further enhancement of the auroral onset arc, and then at \textsc{tha} at $\sim$04:53:20 UT just before auroral poleward expansion. The major $B_z$ increase accompanied by large-amplitude fluctuations in the Pi2 and Pi1 frequency ranges occurred $\sim$1-1.5 min later during poleward expansion. Here, the timing difference between the three spacecraft was due to expansion of the dipolarization region that corresponds to auroral breakup. \textsc{the} further moved toward the equator during dipolarization, as indicated by $|B_x|$ decrease. Dipolarization ended at $\sim$04:56--04:57 UT, corresponding to the end of auroral poleward expansion. The Swarm~A pass at 04:51--04:53~UT (Figure~\ref{fig:synopsis}) was contemporaneous with the onset of the first dipolarization.

Soon after the first dipolarization ceased, dipolarization associated with the second substorm began at \textsc{the} at $\sim$04:58:00 UT at further enhancement of the auroral onset arc, at \textsc{tha} at $\sim$04:58:20 UT just before auroral poleward expansion, and at \textsc{thd} at $\sim$04:58:40 UT just after poleward expansion. Fluctuations associated with the second dipolarization were much larger than those associated with the first dipolarization, probably because the three spacecraft were in the central part of the dipolarization region (see Figure~\ref{fig:synopsis}) and also further approached the equator during the second dipolarization \citep[see ][]{shiokawaMagneticFieldFluctuations2005}. In particular, at $\sim$04:58:30--04:59:30 UT and $\sim$05:00:30--05:02:00 UT, the amplitudes of the fluctuations were extremely large; $B_z$ at \textsc{thd} spiked to ${\sim}+75$~nT, while that at the three spacecraft transiently became even negative. Positive $B_z$ spikes were accompanied by very fast earthward flows exceeding 600 km/s (1000 km/s at \textsc{the}) with large dawnward components exceeding 400 km/s. The second dipolarization ended at $\sim$05:04 UT, with the net $B_z$ increase larger than that of the first dipolarization, corresponding to the greater intensity (auroral activity and geomagnetic negative bay) of the second substorm.

\begin{figure*}
\includegraphics[width=.96\textwidth]{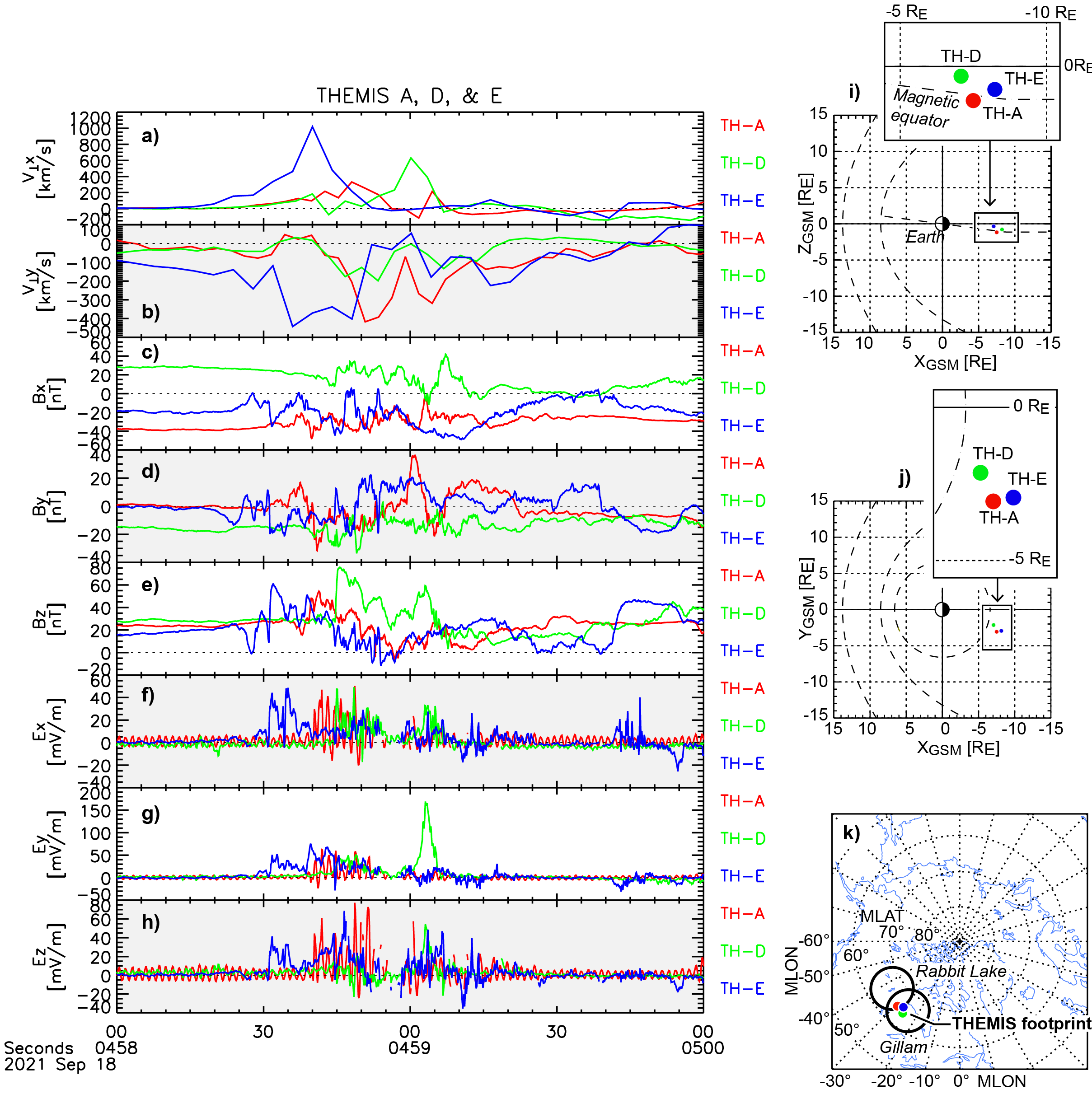}
    \caption{\label{fig:themis} The $X$ and $Y$ components of the ion velocity perpendicular to the magnetic field \textbf{(panels a, b)} and the three components of the magnetic \textbf{(c--e)} and electric \textbf{(f--h)} fields in GSM coordinates from the \textsc{themis} A (red), D (green), and E (blue) spacecraft from 04:58~UT to 05:00~UT on 18 September 2021, along with the locations of the three spacecraft in Earth's magnetotail and their ionospheric footprints \textbf{(panels i--k)}.}
\end{figure*}

As shown below, we identify $\sim$04:58:30--04:59:30~UT as the interval during
which the Alfv\'{e}nic source amplitude maximized and the mapped field at the auroral field-line's E-region foot exceeded the Farley--Buneman threshold in strength.  Figure~\ref{fig:themis}a--h) shows a zoom-in of the magnetic field from the three spacecraft for 04:58--05:00~UT, together with the $X$ and $Y$ components of the ion velocity perpendicular to $\mathbf{B}$ and the three components of the electric field. During this interval of large-amplitude magnetic fluctuations and fast flows, all three spacecraft recorded strongly fluctuating electric fields, directed mainly earthward, duskward, and equatorward, exceeding $20~\mathrm{mV\,m^{-1}}$ and sometimes even $50~\mathrm{mV\,m^{-1}}$. The equatorward $E_z$ reversed sign across the current sheet, positive at \textsc{tha} and \textsc{the}, which were on its southern side ($B_x<0$), and negative at \textsc{thd} on the northern side ($B_x>0$). This reversal is the signature expected of a space-charge (Hall) electric field due to the thin current sheet, as reported by
\cite{luHallElectricField2019}.

Large-amplitude fluctuations, including the equatorward $E_z$, appeared only during the auroral poleward expansion of the second substorm, while they were weak for the first (Figures~\ref{fig:extthem}, \ref{fig:extthem2}, and S1--S3 in the Supplementary Information). This is plausibly because during the first substorm the \textsc{themis} spacecraft were not in the central part of the dipolarization region, their footprints lay outside the main auroral breakup region, and they therefore sampled the plasma sheet away from the equator, as indicated by larger $|B_x|$ and smaller ion $\beta$. After the dipolarization ($\sim$05:04~UT) the current sheet thickened and the ions re-magnetised, and we conjecture that the local space-charge field subsequently decayed.

We next relate the near-equatorial fields to the ionospheric foot. The three spacecraft encountered similar structured fields in sequence (Figure~\ref{fig:themis}d); combined with their relative positions (panels i and j), the front motion (expansion of the dipolarization region) has an earthward speed ($\sim$200~$\mathrm{km\,s^{-1}}$, from the radial ordering of the dipolarization onsets) and a comparable dawnward speed ($\sim$400~$\mathrm{km\,s^{-1}}$). A single propagation vector is underconstrained with three spacecraft, so we treat these as components of the front motion rather than a unique velocity. Independently, the \textsc{themis} ion moments give an earthward bulk velocity of $\sim$300--500~$\mathrm{km\,s^{-1}}$ (Figure~\ref{fig:themis}a) with a dawnward component of similar magnitude
(panel b).

Three physically distinct contributions enter the source-side perpendicular electric field, and they map to the ionosphere differently. Two are genuine electric fields that enter the wave-driving budget,
\begin{equation}
    \mathbf{E}_{\perp,0} \;=\; \mathbf{E}^{\rm conv} + \mathbf{E}^{\rm dipol},
\end{equation}
while the third is a ``purely kinematic pattern'', or, rather, a phase velocity, treated separately below.

\textit{(i) Convective field of the background flow.} In the frozen-in limit $\mathbf{E}^{\rm conv} = -\mathbf{v}_i\times\mathbf{B}_{\rm post\text{-}dipol}$. With the bulk flow essentially in the equatorial plane and the post-dipolarization field nearly along the dipole axis, the magnitude reduces to $|\mathbf{E}^{\rm conv}|\approx v_i\,B_z \approx (400~\mathrm{km\,s^{-1}})(25\text{--}40~\mathrm{nT}) \approx 10\text{--}15~\mathrm{mV\,m^{-1}}$, pointing duskward in GSM. Mapped along the field line this delivers a westward $\mathbf{E}$ at the foot and an \emph{equatorward}  $\mathbf{E}\times\mathbf{B}$ drift.

\begin{table}[t]
\centering
\caption{Summary of the per-spacecraft WKB amplitude maps along the \textsc{themis} field lines, using the McIlwain $L$ from the T96 external field at 04:58~UT 2021-09-18. $\lambda_f$ is the dipole-equivalent foot invariant latitude; $B_0$ and $v_{A,0}$ are the equatorial (source) field strength and Alfv\'{e}n speed, respectively; $\mathrm{amp_{top}}$ and $\mathrm{amp_{foot}}$ are $E_\perp/E_{\perp,0}$ at the topside $v_A$ peak and at the MHD cutoff (the latter is the operative, pre-reflection value); $\tau$ is the source$\to$foot Alfv\'{e}n transit time. The mapped foot field is $E_\perp^{\rm foot}=T\,\mathrm{amp_{foot}}\,E_{\perp,0}$ with $E_{\perp,0}=20$--$50~\mathrm{mV\,m^{-1}}$ (Eq.~\ref{eq:E0}) and $T=0.16$--$0.60$ ($\Sigma_P=2$--$10$~S, $\Sigma_A\approx0.85$~S). Values are from the simulation runs of Fig.~\ref{fig:simulations}, described in Appendix~A.}
\label{tab:wkb}
\begin{tabular}{lcccccccc}
\hline
Probe & $L_{\rm T96}$ & $\lambda_f$ & $B_0$ & $v_{A,0}$ & $\mathrm{amp_{top}}$ & $\mathrm{amp_{foot}}$ & $\tau$ & $E_\perp^{\rm foot}$ \\
      &               & [deg]       & [nT]  & [$\mathrm{km\,s^{-1}}$] &           &            & [s]   & [$\mathrm{mV\,m^{-1}}$] \\
\hline
\textsc{thd} & 8.8  & 70.3 & 46 & 1900 & 36.7 & 24.5 & 16.3 & 78--735 \\
\textsc{tha} & 10.0 & 71.6 & 31 & 1200 & 55.4 & 36.2 & 26.0 & 116--1086 \\
\textsc{the} & 11.6 & 72.9 & 20 &  800 & 89.5 & 56.9 & 45.4 & 182--1707 \\
\hline
\end{tabular}
\end{table}

\textit{(ii) Dipolarization-associated space-charge (Hall) field.} The equatorward $E_z$ identified above is the space-charge field during dipolarization and is the dominant perpendicular field during the activation, exceeding $50~\mathrm{mV\,m^{-1}}$ in the bursts; its coherent, band-limited (Pi1) amplitude is several tens of $\mathrm{mV\,m^{-1}}$. Mapped to the Northern-Hemisphere foot, an equatorward $\mathbf{E}$ crossed with the (downward) ionospheric $\mathbf{B}$ yields an \emph{eastward} $\mathbf{E}\times\mathbf{B}$ drift. After the WKB amplification described below this is the field that drives the eastward E-region irregularity drift  $>5000~\mathrm{m\,s^{-1}}$ observed by \textsc{icebear}, which requires a mapped foot field $>$250~$\mathrm{mV\,m^{-1}}$, well above the Farley--Buneman threshold ($E_\perp\gtrsim 20~\mathrm{mV\,m^{-1}}$,  i.e.\ a drift exceeding the ion-acoustic speed). The dipolarization-associated space-charge field is therefore the driver of the central observable, not the convective term \textit{(i)}.

\textit{(iii) Azimuthal phase propagation of the dipolarization.} If the dawnward propagation of structured fields (Figure~\ref{fig:themis}d) exceeds the dawnward \emph{bulk}-flow speed from the ion moments, the excess is a phase velocity, or the rate at which the dipolarization activates successive flux tubes (the azimuthal expansion of the substorm), not a bulk drift of a single tube. It is not a field-line $\mathbf{E}\times\mathbf{B}$ velocity \citep{newcombMotionMagneticLines1958}, and so it is not an additional $\mathbf{E}\times\mathbf{B}$ drift to add to the budget: the electric field accompanying the structure's passage at any point is the local space-charge field $\mathbf{E}^{\rm dipol}$ already counted in \textit{(ii)}, and adding $v_{\rm phase}\mathbf{B}$ would double-count the front's $\partial\mathbf{B}/\partial t$ (Appendix~A).

As explained in Appendix~A, for the amplitude budget, we adopt the band-limited perpendicular source amplitude
\begin{equation} \label{eq:E0}
    |\mathbf{E}_{\perp,0}| \;=\; |\mathbf{E}^{\rm conv}+\mathbf{E}^{\rm dipol}|
    \;\sim\; 20\text{--}50~\mathrm{mV\,m^{-1}},
\end{equation}
consistent with the measured perpendicular fields (Figures~\ref{fig:extthem}--\ref{fig:themis}) and dominated by the Hall
term \textit{(ii)}.

We map $E_\perp$ to the ionospheric foot with the slowly-varying-envelope WKB calculation of Appendix~A, run separately along each spacecraft's field line. We adopt the McIlwain $L$ evaluated with the T96 external field at 04:58~UT ($L=8.8$, $10.0$, and $11.6$ for \textsc{thd}, \textsc{tha}, and \textsc{the}), which places the dipole-equivalent feet at $70.3^\circ$, $71.6^\circ$, and $72.9^\circ$ invariant latitude. These feet lie a few degrees poleward of the Gillam zenith; the TREx imagery shows a continuous, quiescent auroral arc spanning the two theatres with the activation on its poleward side, so that the \textsc{themis} tubes thread the source region while \textsc{icebear}, a few degrees equatorward and to the west, samples the E-region response within the same structure. The three maps are summarised in Table~\ref{tab:wkb}. The pre-reflection foot amplification is $E_\perp/E_{\perp,0}\approx 24$--$57$, and the source$\to$foot Alfv\'{e}n transit time is $16$--$45$~s. For the most equator-conjugate probe (\textsc{thd}) the transit is $16$~s, and so a pulse launched at the $\sim$04:58:30~UT activation reaches the foot well within the identified 04:58:30--04:59:30~UT interval, and remains distinctly
Alfv\'{e}nic (seconds) relative to the minutes expected of slower magnetosonic or drift communication.

After the thin-sheet reflection  ($T=2\Sigma_A/(\Sigma_A+\Sigma_P) =0.16$--$0.60$ for $\Sigma_P=2$--$10$~S), the mapped foot field is $E_\perp^{\rm foot}=T\,(E_\perp/E_{\perp,0})\,E_{\perp,0}$, which for the most equator-conjugate probe (\textsc{thd}) and a representative $\Sigma_P$ reproduces the extreme $\sim$330~$\mathrm{mV\,m^{-1}}$ inferred from the \textsc{icebear} drifts (see Figure~\ref{fig:ice2} in the next subsection); the full envelope (Table~\ref{tab:wkb}) comfortably brackets the observations. We therefore read the budget as a demonstration that the observed foot field is energetically attainable \textit{from the measured source}. 


\subsection{E-region coherent scatter and target motions}

\begin{figure}
    \centering
    \includegraphics[width=1.1\textwidth]{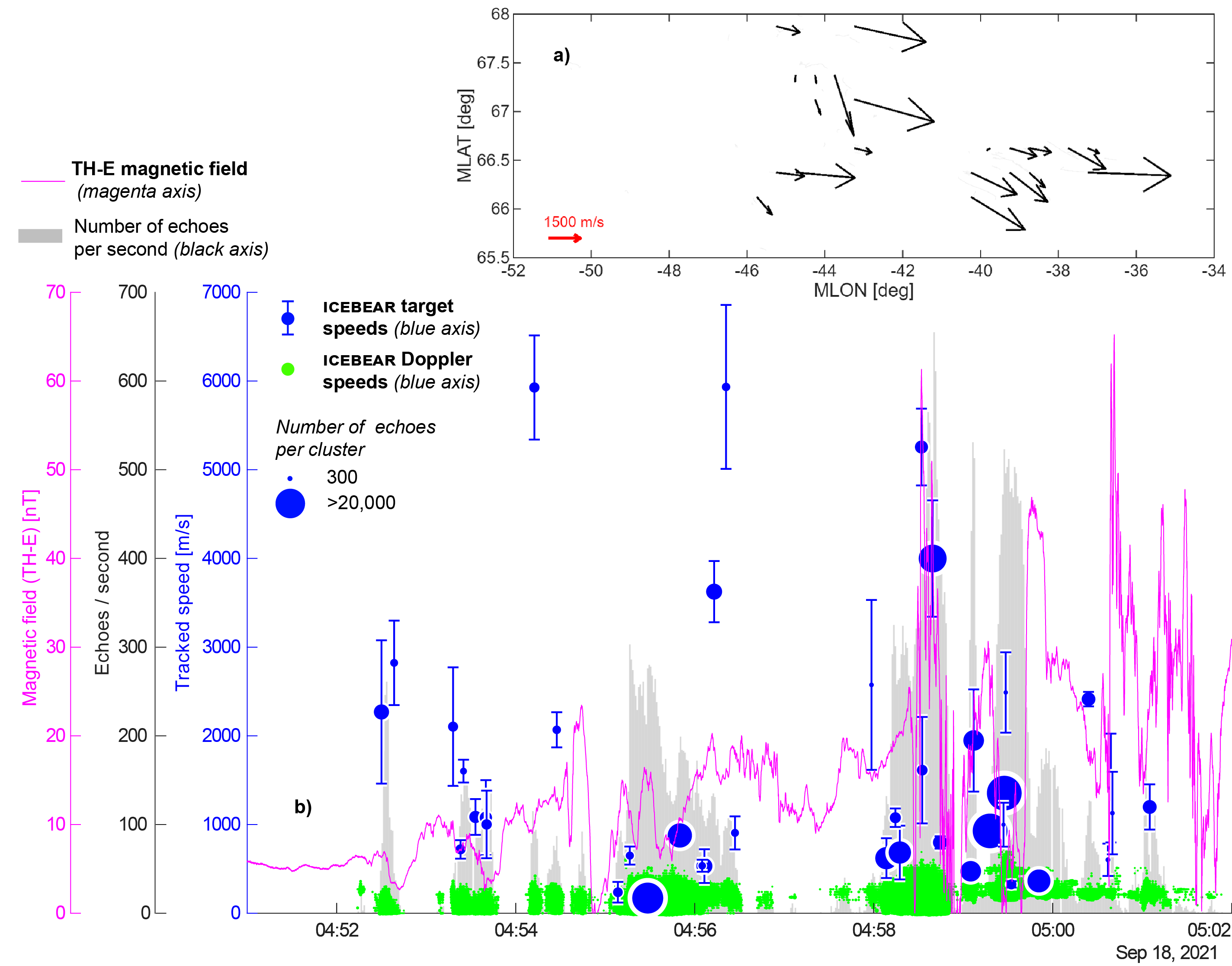}
\caption{
\textbf{Panel a)} shows a spatial and temporal average vector field (showing observed radar target motions), constructed from data  observed between 04:57:30~UT and 05:01:00~UT. We use altitude-adjusted corrected geomagnetic (AACGM) coordinates \citep{bakerNewMagneticCoordinate1989}.  \textbf{Panel b)} shows the radar echo detection rate for this interval (shaded region, black $y$-axis),  the tracked radar speeds (blue circles, blue $y$-axis) and Doppler speeds (green circles, blue $y$-axis). Blue errorbars denote 95-percentile confidence intervals for  the velocity estimation, and the circles' sizes reflect number of echoes per tracked cluster. Superposed, we show the $z$-component of the magnetic field observed \textit{in-situ} by \textsc{themis}-E (magneta line, magenta $y$-axis). See Figure~\ref{fig:swarm} for the Swarm~A conjunction.}
    \label{fig:bulkmotions}
\end{figure}

Figure~\ref{fig:bulkmotions} shows coherent scatter observations from \textsc{icebear}, using the recently improved point-cloud analysis techniques originally due to \cite{ivarsen_point-cloud_2024}. Starting from around 04:54~UT, \textsc{icebear} detected discrete clusters of E-region backscatter exhibiting unusually high eastward target velocities on the \textit{poleward edge} (see Figure~\ref{fig:synopsis}) of the developing auroral forms. Figure~\ref{fig:bulkmotions}a) shows the spatially and temporally averaged spatial tracks, as a collection of black arrows, of these clusters, while Figure~\ref{fig:bulkmotions}b) shows the tracked speeds and their timing, with the \textsc{themis}-E magnetic field ($B_z$) data overlaid.

We note a significant discrepancy between these target motions and the internal Doppler velocities of the echoes (green data in Figure~\ref{fig:bulkmotions}b). While the target displacements reached $\sim$6000 m/s, the corresponding phase velocities of the waves themselves (that is, the radar Doppler speeds), remained $<500$~m/s. This is consistent with established instability theory, where Farley-Buneman wave phase velocities are typically bounded by the local ion acoustic speed \citep{sahr_auroral_1996,fosterSimultaneousObservationsEregion2000,oppenheim_kinetic_2013,chauUnusualRegionFieldaligned2016}. Because the tracked motion of the echo cluster tracks the displacement of the instability source region rather than the individual waves, the tracks become proxy measurements for local ionospheric enhancements in the $\mathbf{E}\times\mathbf{B}$ drift speed, and thus the ionospheric electric field \citep{ivarsen_point-cloud_2024,ivarsen_deriving_2024}.

The appearance of these super-high-velocity structures is both spatially and temporally constrained. They emerge  during the 04:54–05:00~UT window, matching the timeline of the magnetic field reconfiguration (dipolarization) observed by the \textsc{themis} spacecraft in the magnetotail (magenta line in Figure \ref{fig:bulkmotions}b).

We suggest a mechanism modulating the local E-region electric field by sudden intense Alfv\'{e}n wave trains excited during the tail current disruption. This Alfv\'{e}nic modulation interacts constructively with localized conductivity-gradients on the edges of the auroral arc \citep[see, e.g., ][]{fujii_reformulation_2011}, and the composite field provides a plausible explanation for the transient, intense electric field structures required to drive the high-velocity E-region radar target motions observed by \textsc{icebear}.

\begin{figure}
    \centering
    \includegraphics[width=1.1\textwidth]{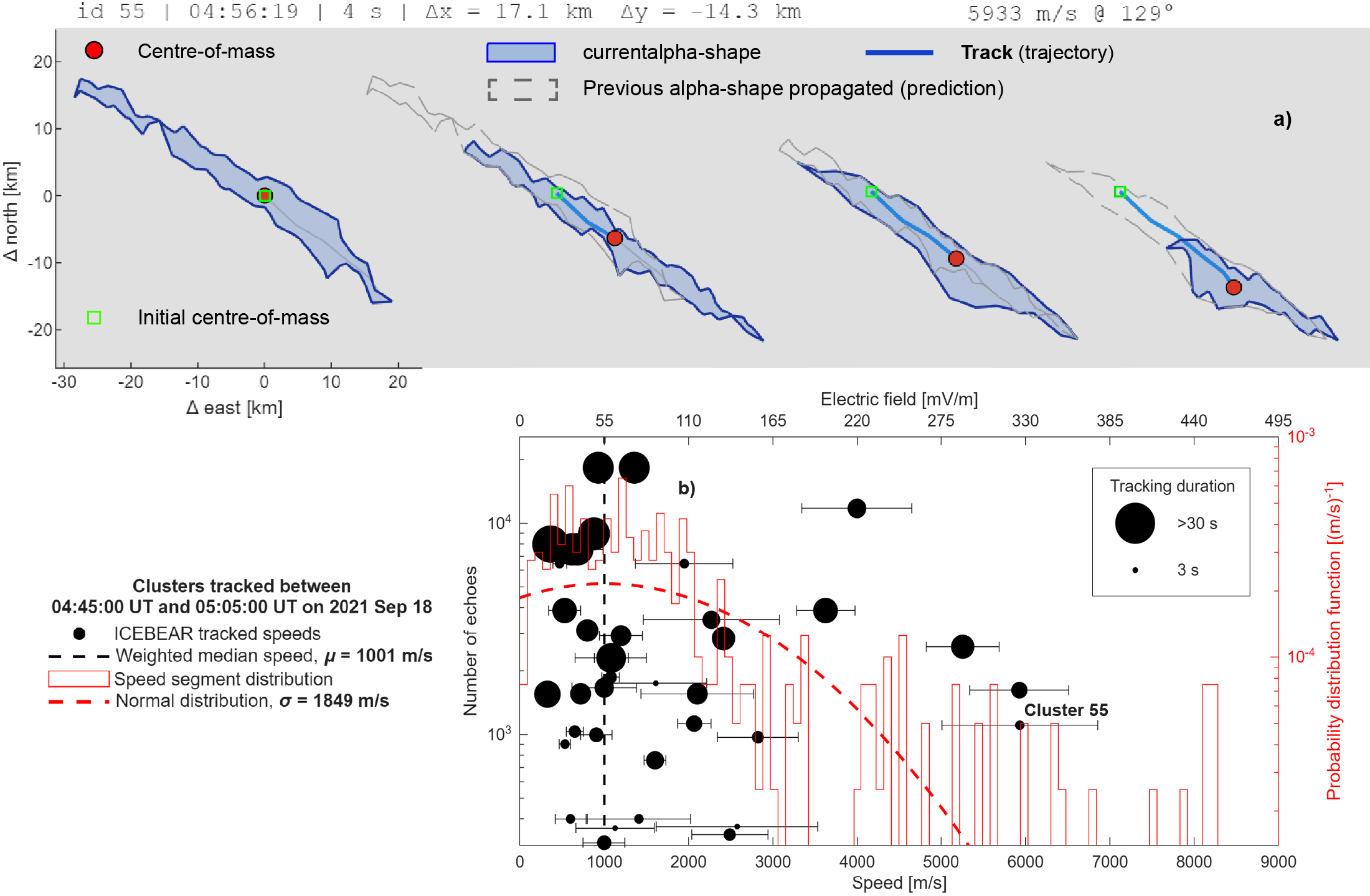}
    \caption{\textbf{Panel a):} The evolution and trajectory (time-history) of `Cluster 55', plotted akin to Figure~\ref{fig:icebear}, with the south-eastward velocity of 5933~m/s indicated by the linear regression. \textbf{Panel b)} shows the distribution of tracked velocities; this is plotted by number of echoes per track against speed (black circles with horizontal errorbars, left $y$-axis), and the sizes of each circle scale with the tracking duration for that cluster. The weighted median of 1011~m/s is shown with a dashed, black line. The distribution of underlying segment-based linear regressions are shown with a red histogram (right $y$-axis). `Cluster 55', the subject of panel a), is indicated in panel b). }
    \label{fig:ice2}
\end{figure}

Before describing the \textit{in-situ} observations by Swarm~A, which occurred near the beginning of dipolarization associated with the first substorm, we must dwell for a moment on the extreme nature of the tracking radar observations. Figure~\ref{fig:ice2}a) shows the evolution of a particularly fast field-structure, where the echo centre-of-mass inside the cluster rapidly shifted in the south-east direction, exhibiting a speed of $5933$~m/s, likely the fastest-moving radar observation to this date. 

Figure~\ref{fig:ice2}b) aggregates the tracking observations from a 20-minute interval encompassing the event under study, and we fit a one-sided Gaussian curve ($\mu=1001$~m/s, $\sigma=1849$~m/s) to the right-skewed speed distribution. The long tail of this distribution (and the relatively few echoes found to move at such speeds) is consistent with expectations for the overall turbulent structuring of ionospheric electric fields, and the short tracking duration for these super-fast structures  \citep[usually $<10$~s, ][]{ivarsen_eastward_2025-1} is consistent with strong intermittency. The lack of such super-fast observations prior to \textsc{icebear} is likely due to that radar's excellent measurement fidelity; a conventional SuperDARN radar operates with a temporal cadence that is \textit{60 times longer}.

\subsection{F-Region electrodynamics: Swarm A observations}

\begin{figure*}
    \centering
    \includegraphics[width=1.06\textwidth]{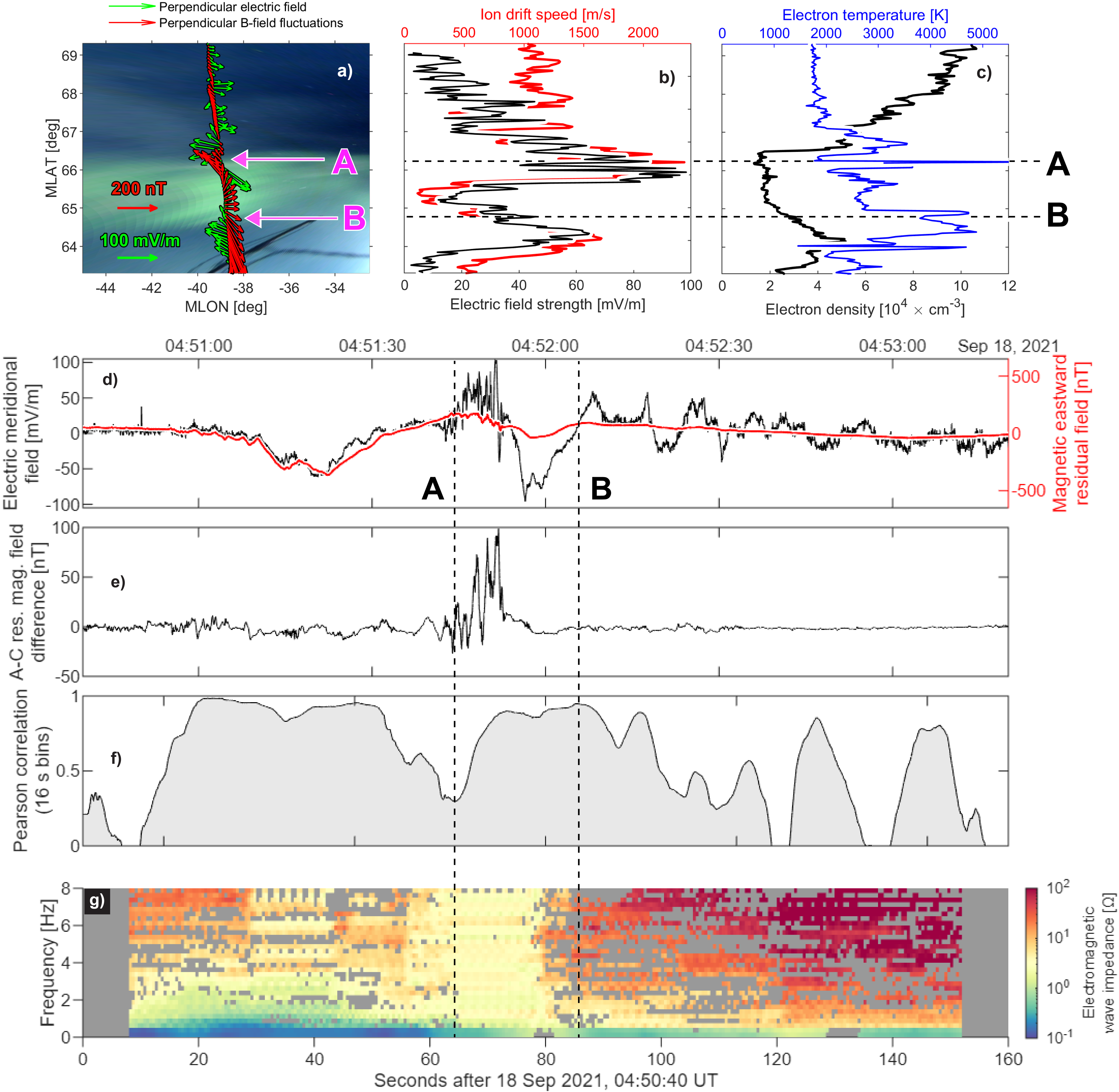}
    \caption{Panel a) shows Swarm A's orbital trajectory superposed on a concurrent auroral image taken at the Gillam station, displaying perpendicular electric field (green arrows) and magnetic field fluctuation (red arrows) vectors, in the mean-field-aligned coordinate system \citep[decomposed into magnetic field-parallel, east, and meridional directions, ][]{ivarsenObservationalEvidenceRole2020}. Panel b) shows ion drift (top, red) and electric field (black) magnitudes, and panel c) shows electron temperature (top, blue) and electron density (black). Panel d) compares the meridional electric field component (black) with the eastward magnetic residual (red, right axis), panel e) shows the difference in magnetic residual observed by Swarm~A and C (shifted using a cross-correlation analysis), and panels f) and g) show the Pearson correlation and an impedance spectrogram, respectively, both within running 16-s windows, and the latter in accordance with \cite{pakhotinDiagnosingRoleAlfven2018}'s method of decoherence masking.
     }
    \label{fig:swarm}
\end{figure*}

To bolster the inferences made in the foregoing, in Figure~\ref{fig:swarm}, we show \textit{in-situ} observations by low-Earth orbit-satellite Swarm~A, of the auroral onset (initial brightening) arc that later participated in auroral breakup. At approximately 04:51-–04:53~UT Swarm~A captured the onset of the \textsc{themis}-observed tail dipolarization, as the Swarm-A spacecraft intersected the auroral oval at an altitude of $\sim$450 km.

Figure~\ref{fig:swarm}a--c) details the \textit{in-situ} plasma and electrodynamic measurements during this crossing. Utilizing concurrent ground-based optical imaging, we identify the poleward (location `A') and equatorward (`B') boundaries of the primary visible arc photographed with the \textsc{tre}x \textsc{rgb} system; note, though, that the pixels map near the north-westward horizon seen from Gillam, and that some uncertainty in the mapping is expected [see \cite{gilliesApparentMotionSTEVE2020} and Figure~2d in \cite{ivarsen_turbulence_2024}]. The Swarm measurements demonstrate that the electrodynamic response within the F-region is highly non-uniform across this structure; while the interior core of the arc exhibited relatively stable transverse fields, the boundaries exhibited sharp gradients in plasma density and elevated electron temperatures (Figure~\ref{fig:swarm}c).

The auroral arc was around 130~km thick, making it a candidate for a quiescent, discrete auroral arc \citep{borovskyQuiescentDiscreteAuroral2019,lysakQuietDiscreteAuroral2020}, though this initial brightening arc already began to grow gradually (Figure~\ref{fig:extgillam}).

\begin{figure}
    \centering
    \includegraphics[width=0.84\textwidth]{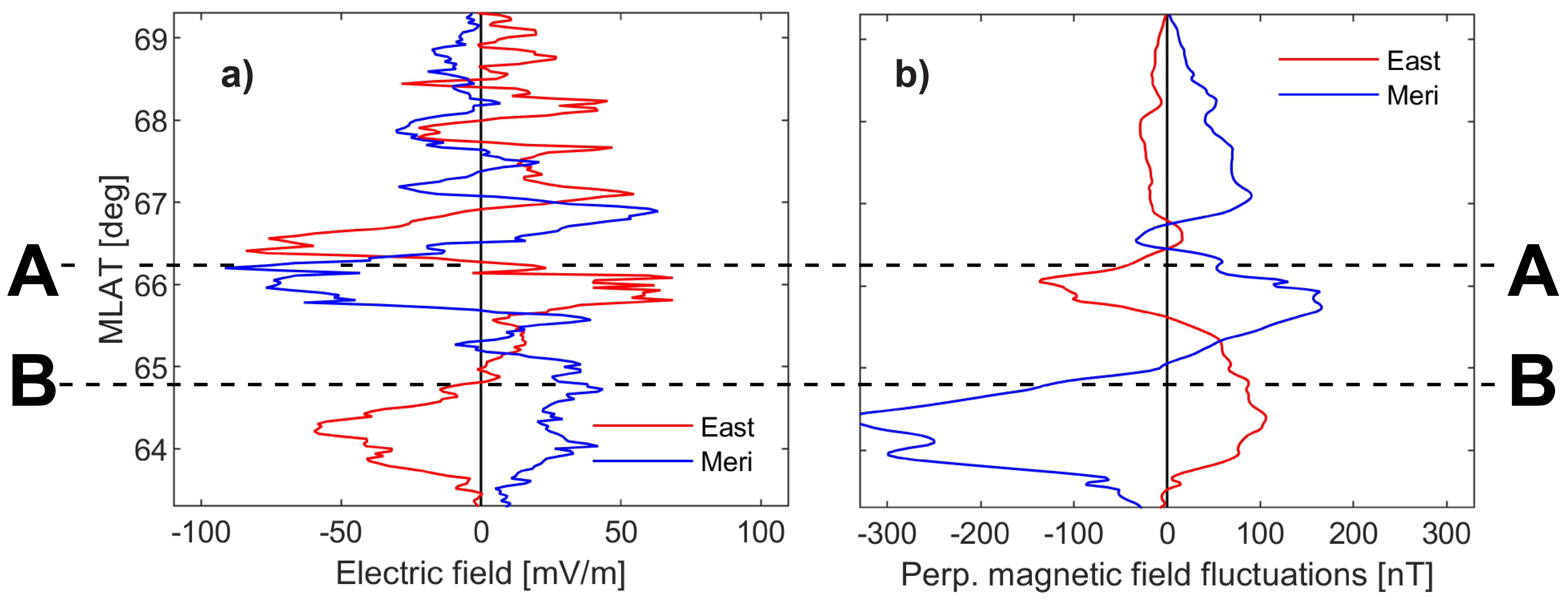}
    \caption{The signed eastward (red) and meridional (blue) components of the electric (panel a) and residual magnetic (panel b) fields, plotted by the satellite's magnetic latitude ($y$-axes), and with the locations `A' and `B' referring to the poleward and equatorward edges of the auroral arc, respectively (see Figure~\ref{fig:swarm}a). Note that panel b) plots magnetic field data at 2~Hz cadence. }
    \label{fig:swarm2}
\end{figure}

Figure~\ref{fig:swarm2} shows the signed eastward (red) and meridional (blue) components of the perpendicular electric (panel a) and residual magnetic (panel b) fields across the pass, plotted against magnetic latitude; we adopt Taylor's hypothesis of frozen-in plasma \citep{fredricksAmbiguitiesDeductionRest1976} to convert the satellite-frame time series to spatial structure. The boundary structure is asymmetric. At `A', the eastward electric field component reverses sharply -- from ${\sim}+50$~mV/m inside the arc to ${\sim}-80$~mV/m just poleward -- but the perpendicular magnetic field shows no coincident transition. At `B', by contrast, $E_\mathrm{East}$ and $\delta B_\mathrm{Meri}$ reverse coincidently, signalling a localized current sheet at that boundary. The largest-amplitude signed structure in the pass, however, is a separate feature located ${\sim}1^\circ$ equatorward of B, near $64^\circ$~MLAT, where $\delta B_\mathrm{Meri}$ executes a ${\sim}500$~nT swing (from $+200$ to $-300$~nT) while $E_\mathrm{East}$ holds at $-50$ to $-75$~mV/m over a comparable latitudinal extent. We interpret this equatorward feature as a distinct, larger-scale electrodynamic structure on adjacent flux tubes, separate from the arc-edge response at `A' and `B', and leave the question of whether it carries primarily Alfv\'{e}nic or quasi-static contributions for dedicated future analysis, though we note analyses of asymmetric current structures in the literature \citep{deboerEffectsMesoscaleRegions2010}.

Using a Swarm-based electromagnetic wave impedance analysis due to \cite{pakhotinDiagnosingRoleAlfven2018}, Figure~\ref{fig:swarm}d--g) disambiguates the wave from the quasi-static contributions to this asymmetric structure. Intervals with high $E$-$B$ Pearson correlation ($\rho>0.9$, Figure~\ref{fig:swarm}f) exhibit frequency-dependent impedance spectra (panel g) consistent with Alfv\'{e}n wave incidence, reflection, and interference within the ionospheric Alfv\'{e}n resonator \citep{knudsenDistinguishingAlfvenWaves1990,lysakFeedbackInstabilityIonospheric1991}, demonstrating that Alfv\'{e}n waves were actively threading the flux tubes connected to the E-region radar echoes. In contrast, the impedance flattened and decreased within the discrete arc (between the vertical black, dashed lines), consistent with a signature controlled by the field-aligned acceleration region rather than by local wave processes. Read together with Figure~\ref{fig:swarm2}, this identifies the boundary at location `A' as dominated by wave-mediated field-aligned Poynting flux, whereas the coincident $E$-$\delta B$ reversal at B is consistent with a quasi-static current sheet. Outside of the discrete arc interior, we suppose or speculate a secondary contribution from Hall-conductance coupling between the $E_\perp$ polarizations at the ionospheric boundary \citep{yoshikawaReflectionShearAlfven1996}, and that the sharp plasma density gradients at location `A' were acting as the primary sites for wave-breaking and mode conversion \citep{shiGenerationElectronAcoustic2018}.

Coincident with the wave signatures, the Langmuir probe and ion drift meters recorded narrow spikes in electron temperature and enhancements in ion drift velocity (Figure~\ref{fig:swarm}b, c). We note that the temperature enhancements occur where the electron density is locally depressed, so that a heat-capacity effect, involving fewer electrons absorbing a given heat flux, may contribute to the observed $T_e$ response alongside any genuine increase in the local heating rate. Here, we note that \cite[][Figure 10]{song_topside_2025} reported a similar offset between the peaks of optical aurora and ionospheric electron temperature.

Swarm thus elucidated the state of the F-region in the early phase of initial auroral brightening. We emphasize that observations at 450~km altitude constrains the F-region plasma environment but does not directly determine the height-integrated Pedersen conductance, which is dominated by E-region densities below ${\sim}150$~km. The visible aurora in the optical imagery confirms that particle precipitation is present across the arc, and the E-region conductance may therefore be substantial even where the F-region density is locally depressed. It remains that the F-region electrodynamic measurements by Swarm provides the necessary physical circumstances to understand the observations of super-fast clusters of FB turbulence tracked by \textsc{icebear}, seen along the poleward edge of the developing auroral forms.

\section{Discussion \& summary}

In this study, we present multi-instrument observations of two substorm expansion phases, capturing a series of substorm-associated dipolarization events in Earth's magnetotail at $X \sim -7$ to $-8$ $R_E$, which started a rapid sequence that culminated in intense, small-scale electrojet plasma turbulence in the E-region ionosphere.

The electrodynamic sequence initiated in the near-Earth plasma sheet. At $\sim$04:51~UT, \textsc{themis} began to capture a distinct tail dipolarization along with fast earthward plasma flows, and Figure~\ref{fig:themis} shows a detailed account of these observations during dipolarization-associated large fluctuations ($\sim$04:58:30~UT). Ion demagnetization in the thinned current sheet generated transient space-charge (Hall) electric fields. The sudden release of free energy launched shear Alfv\'{e}n waves downward along the magnetic field lines toward the bottom of the F-region (Figures~\ref{fig:swarm} and  \ref{fig:simulations}). The transverse electric field created locally by the incidence, reflection, or interference of these waves then mapped electrostatically and semi-instantaneously to the E-region, driving \textit{fast}, turbulent Hall drifts in the auroral electrojets.

The carrier of this event from the magnetotail to Earth's upper atmosphere, Alfv\'{e}n waves, were observed by Swarm at the onset of the dipolarization event, and we infer their continued presence throughout the event.

\subsection{Discussion of the Swarm (F-region) observations of Alfv\'{e}n waves}

Figure~\ref{fig:swarm}f) shows the transverse E-B correlation inside 16-s running, overlapping windows, demonstrating at times very strong correlation ($>0.9$) between $E_\perp$ and $\delta B_\perp$. In the low-frequency MHD limit, a finite, fixed-sign $E_\perp/\delta B_\perp$ admittance (inverse of impedance, Figure~\ref{fig:swarm}g) is the observational signature of a propagating Alfv\'{e}n-mode disturbance along the flux tube \citep{knudsenAlfvenWavesStatic1990,keilingGlobalMorphologyWave2003}. We ascertain the presence of  Alfv\'{e}n waves driving a net  Poynting flux with a component directed along the ambient magnetic field \citep{ivarsenObservationalEvidenceRole2020}, whereby the Alfv\'{e}n modes transmit electromagnetic energy and drive field-aligned currents \citep{alfvenExistenceElectromagneticHydrodynamicWaves1942,siscoeSolarSystemMagnetohydrodynamics1983,yoshikawaReflectionShearAlfven1996}. The field-modulation produced by their interaction with the ionosphere provides the natural coupling of the dipolarization-associated large-amplitude fluctuations to the ionosphere   \citep{lysakElectrodynamicCouplingMagnetosphere1990,keilingGlobalMorphologyWave2003}.

Alfv\'{e}n waves are the likely candidate mechanism capable of carrying the dipolarization-associated fluctuations to the field-line's footprint near-instantaneously (Figure~\ref{fig:bulkmotions}b), though we note that compressional (fast-mode) ultra-low frequency (ULF) waves excited during dipolarization can act as an intermediary channel, as ULF waves are known to mode-couple to shear Alfv\'{e}n waves \citep{watersCouplingFastShear2013}.

The coincident timing of the \textsc{icebear} and \textsc{themis} observations (Figure~\ref{fig:bulkmotions}b) does however require a \textit{very short transit time} for any candidate mapping mechanism other than Alfv\'{e}n waves; our wave transmission analysis shows that the latter would need only 16~s to reach the ionosphere. This lends credence, and we ascertain the continued presence of Alfv\'{e}n wave incidence, reflection, or interference throughout the dipolarization event, and we note \cite{babu_plasma_2024}'s inference of ballooning instability during the event.

The Alfv\'{e}n wave trains encountered structured conductivity on the edges of the precipitation region (Figure~\ref{fig:swarm}d), reshaping the impending field structure into the spatially confined, transverse field elucidated by the \textsc{icebear} target motions to occur on the poleward side of the observed auroral forms. 


\subsection{Discussion of the magnetosphere-ionosphere coupling and the tracked radar targets}

As we outlined in Section~\ref{sec:theory}, we suggest that the three distinct, observed phenomena are causally linked: The transient space-charge (Hall) electric field during dipolarization, located at the magnetotail equator 7--8 Earth radii distant (Figure~\ref{fig:themis}), created the source amplitude of an Alfv\'{e}nic pulse propagating earthward along the converging auroral flux tube. Standard WKB propagation amplifies the perpendicular electric field amplitude  \citep[Figure~\ref{fig:simulations}, ][]{lysakElectrodynamicCouplingMagnetosphere1990}, while partial reflection at the ionospheric boundary suppresses this amplification \citep{yoshikawaReflectionShearAlfven1996}. Consistently and reassuringly, we observe relevant magnetospheric electric fields around 20--50~mV/m, derive and model a WKB amplification factor, and observe local electric field enhancements up to 330~mV/m in the E-region (Figure~\ref{fig:bulkmotions}b).

The onset of tail dipolarization and the generation of fast earthward/dawnward flows in the plasma sheet (starting at $\sim$04:58 UT) occurred simultaneously with auroral poleward expansion (auroral breakup) and the onset of the extreme $>5000$~m/s E-region radar target motions measured by \textsc{icebear}. This ties the ion demagnetization at the thinned current sheet,  known to generate transient space-charge electric fields \citep{luHallElectricField2019}, directly to small-scale plasma turbulence in Earth's upper atmosphere, a remarkably intimate and far-reaching coupling.

The tracked radar motions, which were alluded to by \cite{chauUnusualRegionFieldaligned2016}'s observation of scatter regions moving much faster than the ion acoustic speed, entails a fundamental shift from \textit{Doppler radar} to \textit{tracking radar}. This is a novel and interdisciplinary development in space physics, which adapts tested algorithms from industrial and aviation applications. With \textsc{icebear}, the target radar analysis for the E-region coherent scatter tracking has been developed into an operational state for consistent $\mathbf{E}\times \mathbf{B}$ measurements, and  we elucidate and substantiate this claim in a companion paper \citep{ivarsenPredictiveRadarTracking2026}. Already, we point to a growing body of  literature that is proving the method's efficacy  \citep{ivarsen_point-cloud_2024,ivarsen_deriving_2024}, and the \textsc{icebear} tracking radar has notably and very recently observed fast ($>4000$~m/s), transient (6--12~s), eastward motions in the \textit{ionospheric cusp region} \citep[see Figure~4 in ][]{ivarsen_eastward_2025-1}.

The underlying turbulence, the 3~meter-scale Farley-Buneman waves, are excited by the penetrating, \textit{structured} \citep{boldyrev_spectrum_2012,chenKineticAlfvenWave2013,galtier_entanglement_2015,david_k_perp_2019} Alfv\'{e}nic field modulations, and we note recent reports of small-scale auroral plasma turbulence being driven directly by magnetospheric wave activity \citep{shen_magnetospheric_2024,ivarsen_characteristic_2025}.

As for direct support in the literature for the mechanisms we describe, studies have already demonstrated azimuthal (longitudinal) motion of auroral forms \citep{ogasawara_azimuthal_2011} dictated by this coupling, and the auroral source regions are the origin of auroral backscatter \citep[][e.g.]{ivarsen_turbulence_2024}, and  extremely strong ($>300$~mV/m) electric fields have been suggested to account for FB wave-associated dynamics near aurorae in the recent literature \citep{whiterFinescaleDynamicsFragmented2021}. The established connections supports the notion that the extreme eastward motions observed in Figure~\ref{fig:bulkmotions}b) are the local E-region signature of a sudden magnetotail dipolarization event.

\subsection{A note on magnetohydrodynamics}

Regarding the specific physics of our proposed mechanism, we emphasize that, although novel in its comprehensive reach, the causal chain that we have mapped out is squarely within the established characteristics of the magnetosphere-ionosphere (M-I) coupling, in which magnetotail reconfiguration drives ionospheric electrodynamics through field-aligned current closure and conductance structuring at auroral arc boundaries. Our contention is quantitative: the extreme tracked radar motions are difficult to recover from conductivity-gradient enhancement of a steady mapped field alone, but are readily recovered from the WKB transmission of an Alfv\'{e}nic pulse launched from the observed dipolarization.

A quantitative case for the Alfv\'{e}nic-transmission scenario requires a brief accounting of alternative magnetohydrodynamic drivers of localized, extreme $E_\perp$ at auroral arc boundaries. Ballooning instability \citep[e.g., ][]{kalmoniStatisticalCharacterizationGrowth2015}, invoked to explain small-scale auroral beads at substorm onset, operates in the tail current sheet and imprints the ionosphere through bead-associated flow structuring \citep{hosokawaLargeFlowShears2013,gallardo-lacourt_ionospheric_2014}; in the present event, the beads for the second substorm appear at $\sim$04:56~UT (Figures~\ref{fig:synopsis} and \ref{fig:extgillam}), concurrent with the super-fast target motions (Figure~\ref{fig:bulkmotions}b). Here, we note that Kelvin-Helmholtz (KH) structuring at flow-shear boundaries is a candidate explanation for the superfast drifts \citep{spicherProductionIonosphericIrregularities2020}; see also \cite{keskinenNonlinearEvolutionKelvinHelmholtz1988}. The KHI predicts vortical $E_\perp$ signatures with sign-reversal across the shear, at which point we would expect to see counter-flows as the KH instability demands. The tracked clusters in Figure~\ref{fig:bulkmotions}a) are however pointed largely and clearly eastward, echoing the `eastward transients' observed in the dayside ionosphere \citep{ivarsen_eastward_2025,ivarsen_eastward_2025-1}, and, though the Swarm conjunction occurred some five minutes before the most extreme electric field spikes were observed, the Swarm-observed field signature (Figure~\ref{fig:swarm2}) does not show the paired, vortical sign reversal across either arc boundary that a KH structure would imprint.

Pressure-gradient and ballooning modes at substorm onset \citep{voronkov_coupling_1997,chengKineticBallooningInstability1998} generate azimuthally (magnetic longitudinally) periodic arc structuring on characteristic wavelengths of several hundred kilometres and quasi-stationary footprint signatures, and their presence during the present event was inferred by \cite{babu_plasma_2024}, but, in isolation, were unlikely to produce the meter-scale field-structure that we infer in the auroral electrojets. To produce meter-scale structures, we allude to established kinetic Alfv\'{e}n wave (KAW) cascades \citep{david_k_perp_2019,streltsovReflectionAbsorptionAlfvenic2003}.

A complementary magnetohydrodynamic amplification channel worth discussing is the ionospheric feedback instability (IFI), in which a downward-propagating Alfv\'{e}n wave couples to ionospheric density perturbations through the resulting modulation of $\Sigma_P$, providing positive feedback to the wave's perpendicular electric field on transverse scales larger than the local electron inertial length ($\delta_e \sim$ a few hundred meters in the E-region) and preferentially on sharp density gradients \citep{lysakFeedbackInstabilityIonospheric1991,sydorenko_stabilizing_2017,greene_situ_2025}. The Swarm~A measurements at the poleward arc boundary (location `A' in Figure~\ref{fig:swarm}) document precisely such a configuration. \cite{streltsovReflectionAbsorptionAlfvenic2003} further demonstrated that the IFI can cascade wave power to narrower transverse scales at the boundaries of the downward-current region. We do not require the IFI to recover the observed $\sim 330$~mV/m foot field, as the WKB wave-transmission analysis in Appendix~A does so within plausible parameter ranges, but the IFI's presence at the gradient region inferred by Swarm is plausible and would act in the same direction as our proposed mechanism, sharpening the Alfv\'{e}n-driven field on the very gradient where \textsc{icebear} observes the fast tracked motions.

We therefore conclude that the coupling action of Alfv\'{e}n waves present the most \textit{parsimonious account} \citep{ockhamOperaPhilosophicaTheologica1967} of the joint magnetotail-and-ionosphere observations between 04:51 and 05:04~UT, observations that we have meticulously described in the present paper. The Alfv\'{e}nic-transmission scenario that we outline is favoured by the conjoint satisfaction of temporal coincidence with the dipolarization marker (Figure~\ref{fig:bulkmotions}b), spatial localization at the Pedersen-conductance gradient inferred from Swarm (Figure~\ref{fig:swarm}b--d), the identification \citep{pakhotinDiagnosingRoleAlfven2018} of an active Alfv\'{e}nic M-I interaction channel threading the relevant flux tube observed during the dipolarization event (Figure~\ref{fig:swarm}g), as well as WKB calculations of field-amplification (up to $\times50$) and transit time (16~s) during wave transmission (Figure~\ref{fig:simulations}).

\subsection{Concluding words}

In closing, we have presented a multi-instrument conjunction study spanning several interconnected yet disparate systems in geospace. We show that a dipolarization event in the magnetotail is immediately followed by extremely fast $E\times B$-drifts across adjacent field lines to the dipolarization footprint. The novelty in this work lies in the demonstration of the foregoing, and in using \textsc{icebear} as a target tracking radar for E-region coherent scatter regions. Observations of fast-moving echo clusters yield a routine, reliable estimation of transient, extreme ionospheric electric field enhancements \citep{ivarsen_eastward_2025,ivarsen_eastward_2025-1}, and we note that there are very few existing observations of similarly extreme speeds observed by ground-based ionospheric radar \citep[see, e.g.,][]{opgenoorthRegionsStronglyEnhanced1990,ivarsen_deriving_2024}. Such extreme and violent electric field enhancements constitute key observables in the ongoing monitoring of Earth's space environment, where the most energetic events are most detrimental to radio communication, and their observation is prerequisite to accurate estimates of the global energy budget during storms and substorms in geospace.

\newpage

\appendix
\section{The Alfv\'{e}nic channel}

In this Appendix we describe the mapping between the magnetotail dipolarization and the auroral E-region as a one-dimensional WKB amplitude map along a dipole
field line with empirical plasma profiles
\citep{lysakFeedbackInstabilityIonospheric1991,PropagationAlfvenWaves1999} separately along the field line of each \textsc{themis} spacecraft. The results are collected in Table~\ref{tab:wkb}, and we use \textsc{thd} ($=$~\textsc{themis}-D) ($L=8.8$), the most
equator-conjugate probe, as the illustrative case throughout.

For a shear Alfv\'{e}n wave propagating parallel to $\mathbf{B}_0$ in the MHD
regime, the transverse fields are tied by
\begin{equation}
    E_\perp \;=\; v_A\,\delta B_\perp,\qquad v_A = \frac{B_0}{\sqrt{\mu_0\rho}},
\end{equation}
with polarization orthogonal to both $\mathbf{B}_0$ and $\delta\mathbf{B}_\perp$. The transfer of electromagnetic energy,
the field-parallel Poynting flux, is given by,
\begin{equation}
    S_\parallel \;=\; \frac{E_\perp\,\delta B_\perp}{\mu_0}
    \;=\; \frac{E_\perp^{2}}{\mu_0\,v_A},
\end{equation}
and a flux tube carrying constant magnetic flux $\Phi_B=B_0 A$ has cross-section
$A(s)=\Phi_B/B_0(s)$. For a propagating packet along this field-line with slowly varying envelope (the
WKB ordering $k_\parallel^{-1}\ll L_B$ and $k_\parallel^{-1}\ll L_\rho$, with $L_B$ and $L_\rho$ the
field and density gradient scales) and no dissipation, the wave \emph{energy
flux} through the tube cross-section is conserved
\citep{brethertonWavetrainsInhomogeneousMoving1968,lysakFeedbackInstabilityIonospheric1991},
\begin{equation}
    P \;=\; S_\parallel\,A \;=\; \frac{E_\perp^{2}\,A}{\mu_0\,v_A}
    \;=\; \text{const. along the tube},
\end{equation}
which, after substituting $A=\Phi_B/B_0$, yields,
\begin{equation} \label{eq:map}
    E_\perp(s) \;=\; E_\perp(s_0)\,
    \sqrt{\frac{v_A(s)\,B_0(s)}{v_A(s_0)\,B_0(s_0)}}.
\end{equation}
Equation~(\ref{eq:map}) is the conventional amplification of the transverse
electric field along an inhomogeneous flux tube \citep{PropagationAlfvenWaves1999}.
We stress that it is a \emph{local} relation: the amplitude at any point depends
only on $v_A B_0$ there relative to the source, not on the intervening path, allowing us to calculate amplification at the footprint as a quantity fixed by the \textit{cutoff}, the altitude below which the wave is no longer a freely propagating shear mode, and the \textit{source}.

Before we model the mass density along the field line, we note that the dipole field ratio between the equatorial apex and the foot is \citep{waltIntroductionGeomagneticallyTrapped1994},
\begin{equation}
    \frac{B_f}{B_0} \;=\; \cos^{-6}\lambda_f\,\sqrt{1+3\sin^2\lambda_f},
\end{equation}
where $\lambda_f$ is the magnetic (dipole) latitude of the foot of the flux tube.

We next apply two empirical models, yielding three distinct components of the plasma column, evaluated   for the conditions of 18 September 2021, 05:00~UT at $58.2^\circ$N, $257^\circ$E ($F_{10.7}=82$, $F_{10.7a}=77$, $A_p=20$), with $F_{10.7}$ being the 10.7~cm radio flux from the sun ($F_{10.7a}$ being its 81-day average) and $A_p$ being a disturbance index based on the planetary $Kp$ index:

\textit{(i)} below 2000~km altitude, IRI \citep{bilitza_international_2022}
supplies electron density $N_e(h)$ together with the molecular O$^+$, N$^+$, H$^+$, He$^+$, O$_2^+$, and
NO$^+$ fractional composition, from which $\rho(h)=N_e(h)\sum_j f_j m_j$ is
computed on a 5~km grid;

\textit{(ii)} from 2000~km into the plasmatrough, the
\cite{sheeleyEmpiricalPlasmasphereTrough2001} formula
$n(L')=124\,(3/L')^4+36\,(3/L')^{3.5}\cos[(\pi/2)(\mathrm{MLT}-7.7)/12]$,
rescaled multiplicatively to maintain continuity with the IRI topside
\citep[see][]{carpenterISEEWhistlerModel1992};

\textit{(iii)} from the trough to the apex, a log-linear blend in geocentric $r$
to a nominal post-dipolarization plasma sheet, $n=0.3$~cm$^{-3}$,
$\bar m_i=1$~amu (\textsc{themis} ESA; the mean ion mass is assumed, as
\textsc{themis} carries no ion mass spectrometer in this range, and
$\bar m_i=1$~amu is conservative for the amplification).

The field $B_0(s)$ is the analytic dipole at the relevant $L$, with arc length
$ds/d\lambda = L R_E\cos\lambda\,\sqrt{1+3\sin^2\lambda}$. The MHD foot cutoff is set by $\omega=\nu_{\rm in}(h)$, with $\nu_{\rm in}(h)$, ion-neutral collision frequencies, built from an
NRLMSIS~2.0 neutral profile \citep{emmertNRLMSIS20WholeAtmosphere2021} using the
\cite{schunkIonospheresTerrestrialPlanets1980} non-resonant rate coefficients for
O$^+$ on N$_2$, O$_2$, He, and H and the \cite{banksAeronomy2013} resonant term
for O$^+$ on O. For the representative $\omega=1$~rad\,s$^{-1}$ (wave frequencies Pi1/PiB,
$f\approx0.16$~Hz) adopted throughout, the cutoff falls at $h=246$~km ---
essentially independent of $L$, since it is set by collisions
\citep[see][for similar inferences]{pakhotinDiagnosingRoleAlfven2018}.

Inserting $v_A(s)$ and $B_0(s)$ into Eq.~(\ref{eq:map}) and restricting to the
MHD-valid region ($h>h_{\rm cutoff}$), the \textsc{thd} map runs from the equatorial
source ($s=0$, $h\approx49{,}700$~km, $B_0=46$~nT, $v_A\approx1.8\times10^3$~km\,s$^{-1}$,
$E_\perp/E_{\perp,0}=1$) through a topside $v_A$ maximum
($\approx2.2\times10^4$~km\,s$^{-1}$, where $E_\perp/E_{\perp,0}=36.7$) to the
cutoff ($h=246$~km, $v_A\approx930$~km\,s$^{-1}$), at which the pre-reflection
amplification is
\begin{equation} \label{eq:ampfoot}
    \left.\frac{E_\perp}{E_{\perp,0}}\right|_{\rm foot}^{\rm pre} \;=\; 24.5
    \qquad(\text{THD; see Table~\ref{tab:wkb} for \textsc{tha}, \textsc{the}}).
\end{equation}
The per-probe foot amplifications span $24.5$--$56.9$, increasing outward as the
source field weakens (Table~\ref{tab:wkb}).

The thin-sheet reflection at the foot follows the
\cite{yoshikawaReflectionShearAlfven1996} treatment. The Alfv\'{e}n (wave)
conductance at the cutoff is
\begin{equation}
    \Sigma_A \;\equiv\; \frac{1}{\mu_0\,v_A^{\rm cutoff}} \;\approx\; 0.85~\mathrm{S},
\end{equation}
and the nightside Pedersen conductance on this date likely lies in the climatological
range $\Sigma_P\approx2$--$10$~S
\citep[][and references therein]{kaepplerDataDrivenEmpiricalConductance2023,juusola_empirical_2025}.
The chain from source to conducting sheet is three multiplicative steps:

\emph{Step (i): WKB amplification to the cutoff.} The pre-reflection foot field is
the source amplitude times the cutoff amplification,
$E_\perp^{\rm pre}=(E_\perp/E_{\perp,0})_{\rm foot}\,E_{\perp,0}$; for \textsc{thd} and the
adopted $E_{\perp,0}=20$--$50$~mV\,m$^{-1}$ (main text, Eq.~\ref{eq:E0}),
$E_\perp^{\rm pre}\approx 490$--$1230$~mV\,m$^{-1}$.

\emph{Step (ii): transmission to the sheet.} The perpendicular field at the
conducting sheet (incident plus reflected) relative to the incident wave is
\citep{mallinckrodtRelationsTransverseElectric1978}
\begin{equation}
    T \;\equiv\; \frac{2\Sigma_A}{\Sigma_A+\Sigma_P}\;\approx\;0.16\text{--}0.60
\end{equation}
for the climatological $\Sigma_P$ range above; $T$ is the boundary electric field ratio,
not a power transmission.

\emph{Step (iii): transmitted field at the conducting sheet,} the quantity that
drives the transverse currents and the tracked \textsc{icebear} clusters, that is, the irregularity
$\mathbf{E}\times\mathbf{B}$ drift,
\begin{equation} \label{eq:250}
    E_\perp^{\rm foot} \;=\; T\,E_\perp^{\rm pre}
    \;=\; T\left(\frac{E_\perp}{E_{\perp,0}}\right)_{\rm foot} E_{\perp,0}
    \;\approx\; 80\text{--}735~\mathrm{mV\,m^{-1}}\quad(\text{THD}),
\end{equation}
with the full per-spacecraft information in Table~\ref{tab:wkb}. For
$\Sigma_P\gtrsim\Sigma_A$ ($T\lesssim1$) the foot field cannot exceed the cutoff
amplification, $E_\perp^{\rm foot}\le(E_\perp/E_{\perp,0})_{\rm foot}\,E_{\perp,0}$;
this is the operative upper bound, set \emph{at the cutoff} and not at the larger
topside $v_A$-peak value. The \textsc{icebear}-inferred electric fields
(Figure~\ref{fig:ice2}) fall within this envelope, with the \textsc{thd} central estimate
reproducing the $>$250~mV\,m$^{-1}$ that corresponds to the observed
$>5000$~m\,s$^{-1}$ eastward drift.

\subsection*{Numerical simulations}

\begin{figure}
    \centering
    \includegraphics[width=1.1\linewidth]{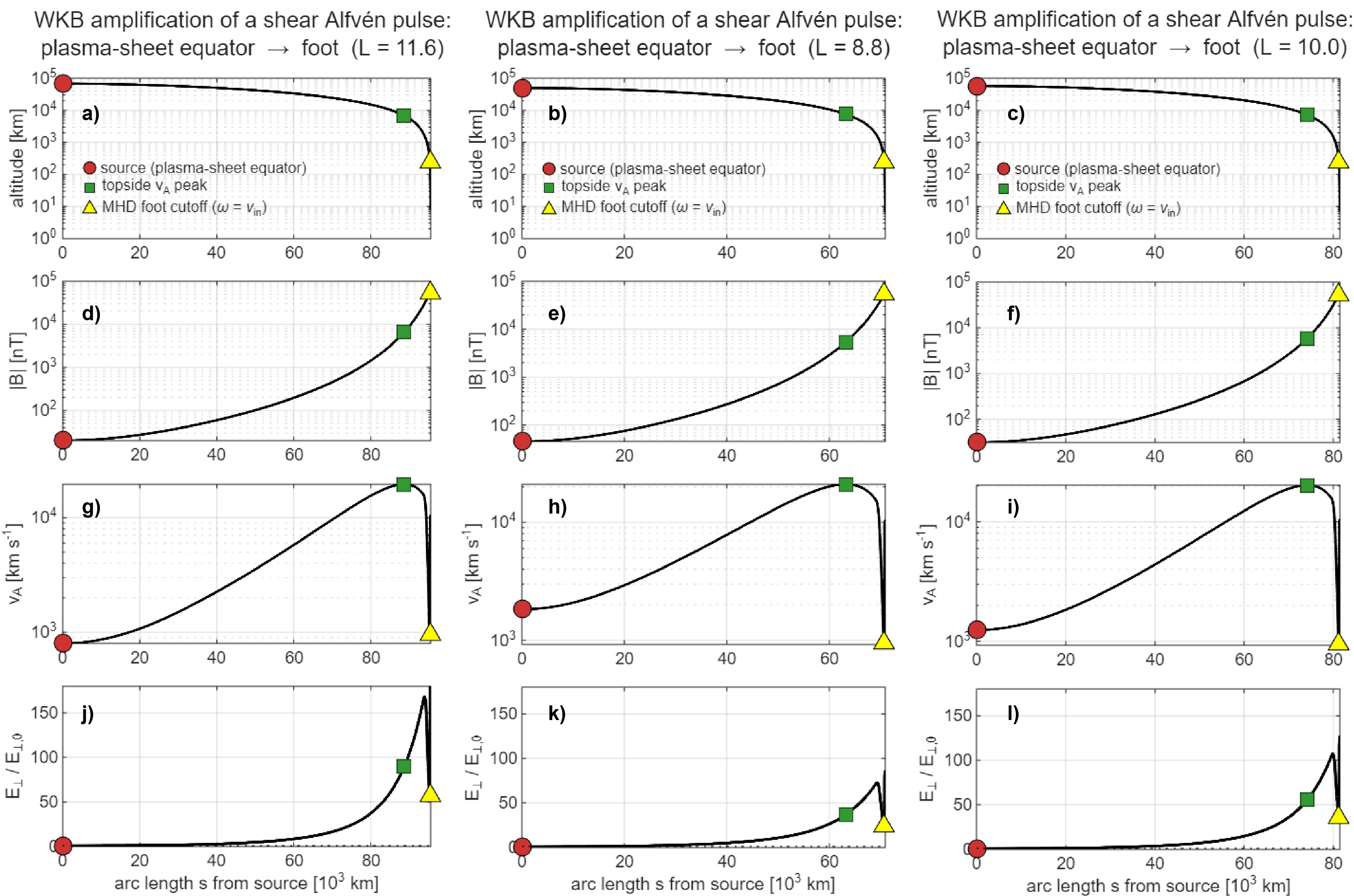}
    \caption{Simulated WKB amplitude maps of a shear Alfv\'{e}n pulse launched at
    the equatorial magnetotail toward the auroral ionosphere, along the three
    \textsc{themis} field lines (T96 McIlwain $L=8.8$, $10.0$, $11.6$ for \textsc{thd},
    \textsc{tha}, \textsc{the}), as functions of arc length $s$ from the source.
    \textbf{Row starting with panel a)} altitude, \textbf{b)} $|B|$, \textbf{c)} $v_A$,
    \textbf{d)} the mapped $E_\perp/E_{\perp,0}$. Markers denote the source
    (equator), the topside $v_A$ peak, and the MHD foot cutoff
    ($\omega=\nu_{\rm in}$). In panel d) the amplitude continues to rise past the
    topside peak to a sharp maximum just above the cutoff; this segment (shaded)
    lies where the WKB ordering fails ($Q\gtrsim1$, see Caveats) and is not a
    physical amplification (the operative value is at the cutoff marker).}
    \label{fig:simulations}
\end{figure}

The empirical mapping above is computed from one-dimensional WKB simulations along
each field-line occupied by a \textsc{themis} spacecraft, shown in Figure~\ref{fig:simulations}.

The amplitude factor (Eq.~\ref{eq:map}) rises from the equatorial source through
the topside $v_A$ maximum and continues to climb past it, because $B$ keeps
rising while $v_A$ has not yet collapsed, and it reaches a sharp maximum just above
the cutoff before $v_A$ falls through the F-region. That near-foot maximum lies in
the region where the WKB ordering fails (see caveats below) and is not physical; the
operative value is at the cutoff (Eq.~\ref{eq:ampfoot}), where the thin-sheet
boundary condition applies.

Numerical integration of $\int_0^{s} \mathrm{d}s'/v_A(s')$ along the same
$v_A(s)$ profile gives a source-to-foot Alfv\'{e}n transit of $16.3$~s for \textsc{thd}
($26.0$ and $45.4$~s for \textsc{tha} and \textsc{the}; Table~\ref{tab:wkb}), dominated by the
low-$v_A$ near-source leg. A pulse launched at the $\sim$04:58:30~UT activation
therefore reaches the 246~km cutoff by $\sim$04:58:46~UT (\textsc{thd}), within the
04:58:30--04:59:30~UT interval; the corresponding uptick in \textsc{icebear} echo
detection (gray histogram, Figure~\ref{fig:bulkmotions}b) makes slower modes of
communicating the dipolarization (magnetosonic, drift-kinetic; minutes) unlikely
candidates for our results.

Lastly, the cutoff at 246~km places the impedance
discontinuity at the base of the O$^+$-dominated F-region rather than in the
E-region Hall layer, consistent with the altitude at which
\cite{pakhotinDiagnosingRoleAlfven2018} place the dominant reflection; this
justifies the thin-sheet approximation, after which the foot field maps
near-instantaneously and electrostatically to electrojet altitudes
($\sim$105~km).

\subsection*{Caveats}

The map is deliberately simple, and fuller treatments nuance the results. The
decisive limitation is that the WKB ordering is weakest where the
amplitude factor is largest: with $\omega=1$~rad\,s$^{-1}$ the parallel wavelength
at the cutoff is $\lambda_\parallel=2\pi v_A/\omega\approx5.8\times10^3$~km, far
exceeding the F-region density scale height, and so the breakdown parameter
$Q\equiv|\mathrm{d}v_A/\mathrm{d}s|/\omega$ exceeds unity over the
topside-peak-to-cutoff leg ($Q\approx2.7$ on average). That $v_A$ gradient is the
upper wall of the ionospheric Alfv\'{e}n resonator
\citep{lysakFeedbackInstabilityIonospheric1991} and it partially reflects the
downgoing pulse, so a full-wave treatment returns a foot amplitude at or below the
WKB value; Eq.~(\ref{eq:ampfoot}) is therefore an upper estimate. Two-dimensional
dipole corrections applied to Pi2 waves
\citep[e.g.][]{streltsovReflectionAbsorptionAlfvenic2003} introduce further
dispersive effects, and we direct the reader to \cite{chaston_turbulent_2008} and
\cite{tianEvidenceAlfvenicPoynting2021} for more complete treatments of this
mapping. For the purposes at hand, the one-dimensional map quantifies the
principle and bounds the amplitude to a sufficient degree.

The displacement current $\varepsilon_0\,\partial\mathbf{E}/\partial t$ is
negligible at the foot. The dimensionless ratio of displacement to Pedersen
conduction current density is
\begin{equation*}
    \frac{|\mathbf{j}_d|}{|\mathbf{j}_P|} \;=\; \frac{\omega\,\varepsilon_0}{\sigma_P},
\end{equation*}
with $\varepsilon_0$ the vacuum permittivity, $\sigma_P$ the local Pedersen
conductivity, and $\omega\approx1$~rad\,s$^{-1}$ the characteristic frequency
(the Pi1/PiB modulation observed by \textsc{themis}). For an auroral E-region peak
$\sigma_P\sim10^{-4}$~S\,m$^{-1}$ this evaluates to $\sim10^{-7}$; equivalently, in
the height-integrated thin-sheet picture, $\varepsilon_0\,\omega\,h/\Sigma_P\sim10^{-7}$
for $h\sim30$~km and $\Sigma_P\gtrsim1$~S. Any inductive
$\partial\mathbf{B}/\partial t$ contribution from the front motion is not an
additional term to budget but is already absorbed into the Alfv\'{e}n-wave
description: Faraday's law $\nabla\times\mathbf{E}=-\partial\mathbf{B}/\partial t$
and the wave-impedance relation $|\mathbf{E}_\perp|=v_A\,|\delta\mathbf{B}_\perp|$
together fix the relation between the wave's $\mathbf{E}_\perp$ and
$\delta\mathbf{B}_\perp$ along the propagating shear mode. Treating the foot field
as the Alfv\'{e}n-wave $\mathbf{E}_\perp$ therefore avoids double-counting.


\section{The interplanetary magnetic field conditions prior to dipolarization}

\begin{figure}
    \centering
    \includegraphics[width=0.6\textwidth]{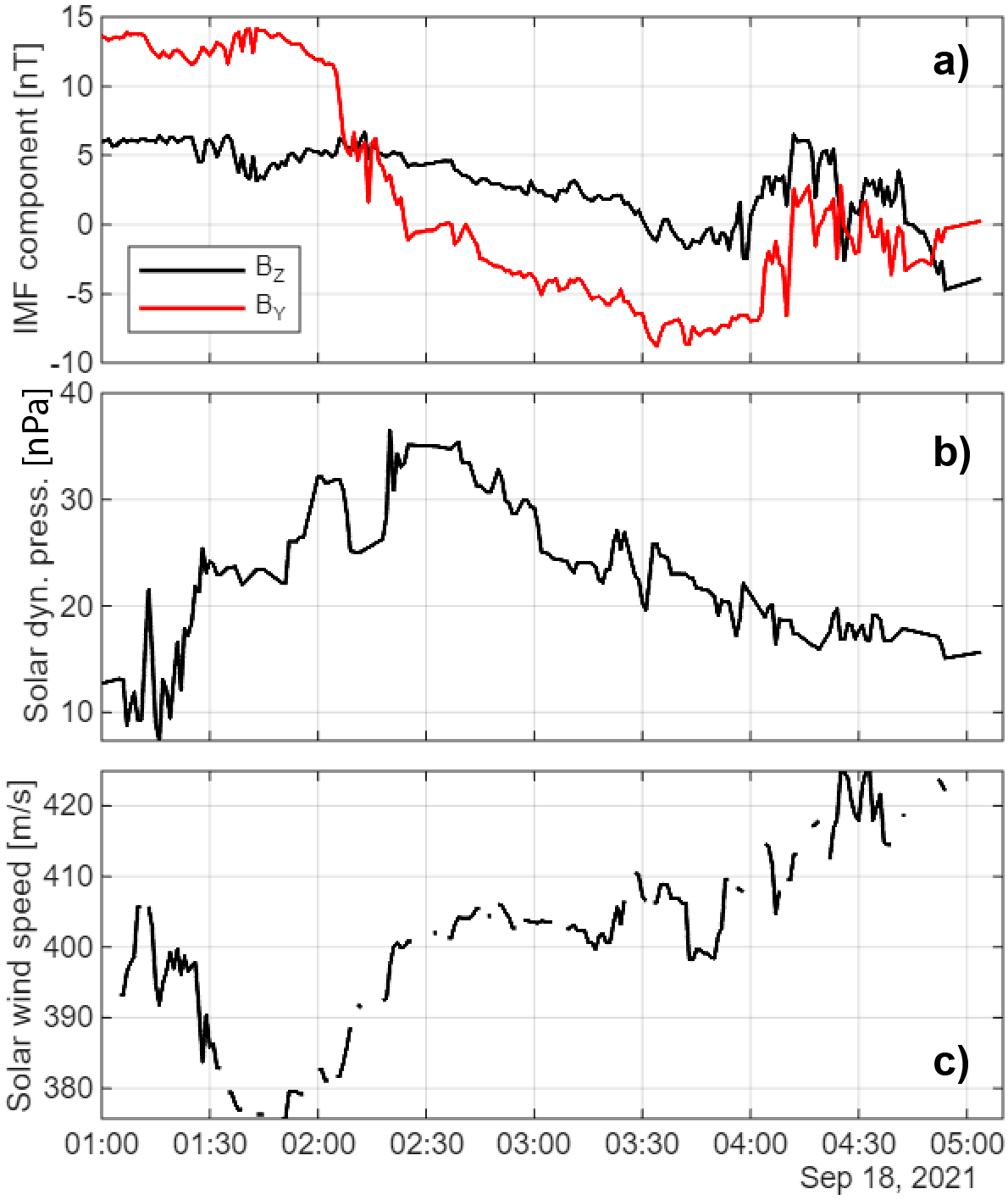}
    \caption{Time-shifted (to the bow shock nose) solar wind- and IMF components from \textsc{omni}. \textbf{Panel a)} shows the IMF $B_Z$ and $B_Y$ components, \textbf{panel b)} shows the solar wind dynamic pressure, while \textbf{panel c)} shows solar wind speed. Note that \textsc{omni} experienced a data gap of several hours following $\sim$05:01~UT.}
    \label{fig:imf}
\end{figure}

Figure~\ref{fig:imf} shows time-shifted \textsc{omni} solar wind conditions from 01:00 to 05:00~UT on 18 September 2021. A compressed sheath arrived near 02:00 UT, marked by an abrupt clock-angle rotation -- $B_Y$ collapsed from +13 to –5~nT in the space of a few minutes, while $B_Z$ began a monotonic decline through zero from its +6~nT pre-event baseline -- and a simultaneous ramp in dynamic pressure to $\sim$35~nPa. The solar wind speed stepped only modestly from $\sim$400 to $\sim$415 km/s (Figure~\ref{fig:imf}c), meaning the pressure enhancement in Figure~\ref{fig:imf}b) was due to enhancements in density rather than an uptick in speed, consistent with a magnetic-cloud sheath or stream-interaction signature rather than a fast-CME shock.

Between 02:15 and 04:00~UT the magnetosphere was exposed to a sustained $B_Y$-dominated IMF ($B_Z \approx 5$~nT $\to$ –2~nT, $B_Y \approx 5$~nT $\to$ –8~nT, with $|B_Y|$ comparable to or exceeding $|B_Z|$), as shown in Figure~\ref{fig:imf}a), while the dynamic pressure remained at 20–30~nPa (Figure~\ref{fig:imf}b). This combination kept dayside closed field line reconnection enabled for $\sim$2.5~h and held the magnetotail energy accumulation throughout the event's growth phase, with the strongly negative $B_Y$ imposing a dawn–dusk asymmetry on the open-flux footprint and the polar-cap convection throat. From 04:00~UT onward $B_Z$ fluctuated back towards zero and $B_Y$ recovered to near zero as the dynamic pressure relaxed toward 15~nPa and the solar wind speed ramped to $\sim$425~km/s (Figure~\ref{fig:imf}c), marking the trailing edge of the sheath \citep{zurbuchenInSituSolarWind2006,kilpuaCoronalMassEjections2017}.

The first substorm expansion onset occurred at $\sim$04:51 UT, which might correspond to a northward turning of the IMF. The 2.5~h interval from southward turning to onset is consistent with the loading window for solar-wind / magnetosphere coupling functions \citep{newellNearlyUniversalSolar2007}, and the timing of expansion onset with respect to the relaxing driver indicates that the external change might have affected the magnetotail internal substorm processes.











\section*{Data availability}

 \textsc{icebear} 3D echo data for 2020-2021 is published with  \url{https://doi.org/10.5281/zenodo.7509022}. \textsc{themis} data can be accessed at \url{https://themis.ssl.berkeley.edu/data/themis/}. Swarm data can be accessed at \url{https://swarm-diss.eo.esa.int}. \textsc{omni} data can be accessed at \url{https://omniweb.gsfc.nasa.gov}.

\section*{Acknowledgements}

This work is supported by the European Space Agency (ESA) Living Planet Grant No. 1000012348 and basic research funding from Korea Astronomy and Space Science Institute (KASI2026183005).
We acknowledge NASA contract NAS5-02099 and V. Angelopoulos for use of data from the \textsc{themis} mission. Specifically, we thank J. P. McFadden and C. W. Carlson for use of the \textsc{themis} ESA data, D. Larson and R. P. Lin for use of the \textsc{themis} SST data, and J. W. Bonnell and F. S. Mozer for use of EFI data. We thank K.-H. Glassmeier, H. U. Auster, and W. Baumjohann for use of the \textsc{themis} FGM data provided under the lead of the Technical University of Braunschweig and with financial support through the German Ministry for Economy and Technology and the German Center for Aviation and Space (DLR) under contract 50 OC 0302.
We thank J. H. King and N. E. Papitashvili for the OMNI solar wind data. Anthropic's Claude Opus 4.8 has been used to assist mathematical formalism, coding in \textsc{matlab}, and information collation.


\end{document}